\pdfoutput=1 
\documentclass{nature}

\usepackage{multirow}
\usepackage{ccaption}
\usepackage{subcaption}
\usepackage{graphicx}
\usepackage{lipsum}
\usepackage{ragged2e}
\usepackage{fixltx2e}
\usepackage{hyperref}

\RequirePackage{mathpazo}

\usepackage{comment}
\usepackage{cite}
\usepackage{adjustbox} 

\makeatletter
\newenvironment{figurehere}
{\def\@captype{figure}}
{}
\makeatletter

\usepackage{xcolor}

\title{\begin{center}  Ferroelectric FET based  Context-Switching FPGA Enabling Dynamic Reconfiguration for Adaptive Deep Learning Machines \end{center}}

\author
{ \centering
{
\large{Yixin Xu$^{1*}$, Zijian Zhao$^{2*}$, Yi Xiao$^{1*}$, Tongguang Yu$^{1}$,  \\Halid Mulaosmanovic$^{3}$,
 Dominik Kleimaier$^{3}$, Stefan Duenkel$^{3}$, \\
 Sven Beyer$^{3}$, Xiao Gong$^{4}$, Rajiv Joshi$^{5}$, X. Sharon Hu$^{6}$, Shixian Wen$^{7}$,  \\
Amanda Sofie Rios$^{7}$, Kiran Lekkala$^{7}$,
Laurent Itti$^{7}$, Eric Homan$^{1}$, \\ Sumitha George$^{8}$,Vijaykrishnan Narayanan$^{1}$, Kai Ni$^{2\dagger}$\\}
\vspace{3ex}
\normalsize{$^{1}$Pennsylvania State University, State College, PA 16802, USA}\\
\normalsize{$^{2}$Rochester Institute of Technology, Rochester, NY 14623, USA}\\
\normalsize{$^{3}$GlobalFoundries Fab1 LLC \& Co. KG, Dresden, Germany}\\
\normalsize{$^{4}$National University of Singapore, Singapore 119077}\\
\normalsize{$^{5}$IBM Thomas J.Watson Research Center, NY 10562 USA.}\\
\normalsize{$^{6}$University of Notre Dame, Notre Dame, USA}\\
\normalsize{$^{7}$University of Southern California, Los Angeles, CA 90089, USA}\\
\normalsize{$^{8}$North Dakota State University, Fargo, ND 58102, USA}\\
\vspace{2ex}
\normalsize{$^{*}$Equally contributing authors} \\
\normalsize{$^{\dagger}$To whom correspondence should be addressed} \\
\normalsize{Email: kai.ni@rit.edu} \\
}}

\begin{document}
\flushbottom
\maketitle
\vspace{3ex}
\begin{abstract}
Field Programmable Gate Array (FPGA) is widely used in acceleration of deep learning applications because of its reconfigurability, flexibility, and fast time-to-market. However, conventional FPGA suffers from the tradeoff between chip area and reconfiguration latency, making efficient FPGA accelerations that require switching between multiple configurations still elusive. In this paper, we perform technology-circuit-architecture co-design to break this tradeoff with no additional area cost and lower power consumption compared with conventional designs while providing dynamic reconfiguration, which can hide the reconfiguration time behind the execution time. Leveraging the intrinsic transistor structure and non-volatility of ferroelectric FET (FeFET), compact FPGA primitives are proposed and experimentally verified, including 1FeFET look-up table (LUT) cell, 1FeFET routing cell for connection blocks (CBs) and switch boxes (SBs). To support dynamic reconfiguration, two local copies of primitives are placed in parallel, which enables loading of arbitrary configuration without interrupting the active configuration execution. A comprehensive evaluation shows that compared with the SRAM-based FPGA, our dynamic reconfiguration design shows 63.0\%/71.1\% reduction in LUT/CB area and 82.7\%/53.6\% reduction in CB/SB power consumption with minimal penalty in the critical path delay (9.6\%). We further implement a Super-Sub network model to show the benefit from the context-switching capability of our design. We also evaluate the timing performance of our design over conventional FPGA in various application scenarios. In one scenario that users switch between two preloaded configurations, our design yields significant time saving by 78.7\% on average. In the other scenario of implementing multiple configurations with dynamic reconfiguration, our design offers time saving of 20.3\% on average.  
\end{abstract}

%
%
\thispagestyle{empty}


\section*{\large Introduction}

Deep neural networks (DNNs) have dominated artificial intelligent (AI) applications due to their cutting edge performance in a wide range of applications in many domains, such as image classification\cite{he2016deep, huang2017densely}, object detection\cite{ren2015faster, redmon2016you}, and natural language processing\cite{devlin2018bert, liu2019roberta}. However, with more sophisticated models and more voluminous data to process \cite{mehonic2022brain}, these DNN workloads are becoming more compute-intensive and data-intensive, requiring hardware accelerators to achieve lower latency, higher throughput, and higher energy efficiency. FPGA devices, with the capabilities of flexible reconfiguration for arbitrary logic functions while maintaining high performance, are gaining popularity as accelerators for such complex deep learning applications\cite{fpga_dcnn_acc, fpga_reconf_nn_acc, fpga_nn_survey}. 
The reconfigurability of FPGA is enabled by its unique architecture, as illustrated in Fig.~\ref{fig:DLN}(a), which consists of a sea of configuration logic blocks (CLBs), CBs, SBs, configuration memory, and I/O blocks\cite{brown1992field}. In particular, CLBs are the main components that can be programmed to perform different logic operations and CBs and SBs are controlled by configuration bits loaded from the configuration memory. A variety of routing networks can be achieved through loading different configuration bits. Above all, FPGA's aforementioned properties including reconfigurability, flexibility, high performance, and fast time-to-market makes it a promising choice for DNN accelerators. 

\begin{figurehere}
   \centering
    \includegraphics[scale=1,width=\textwidth]{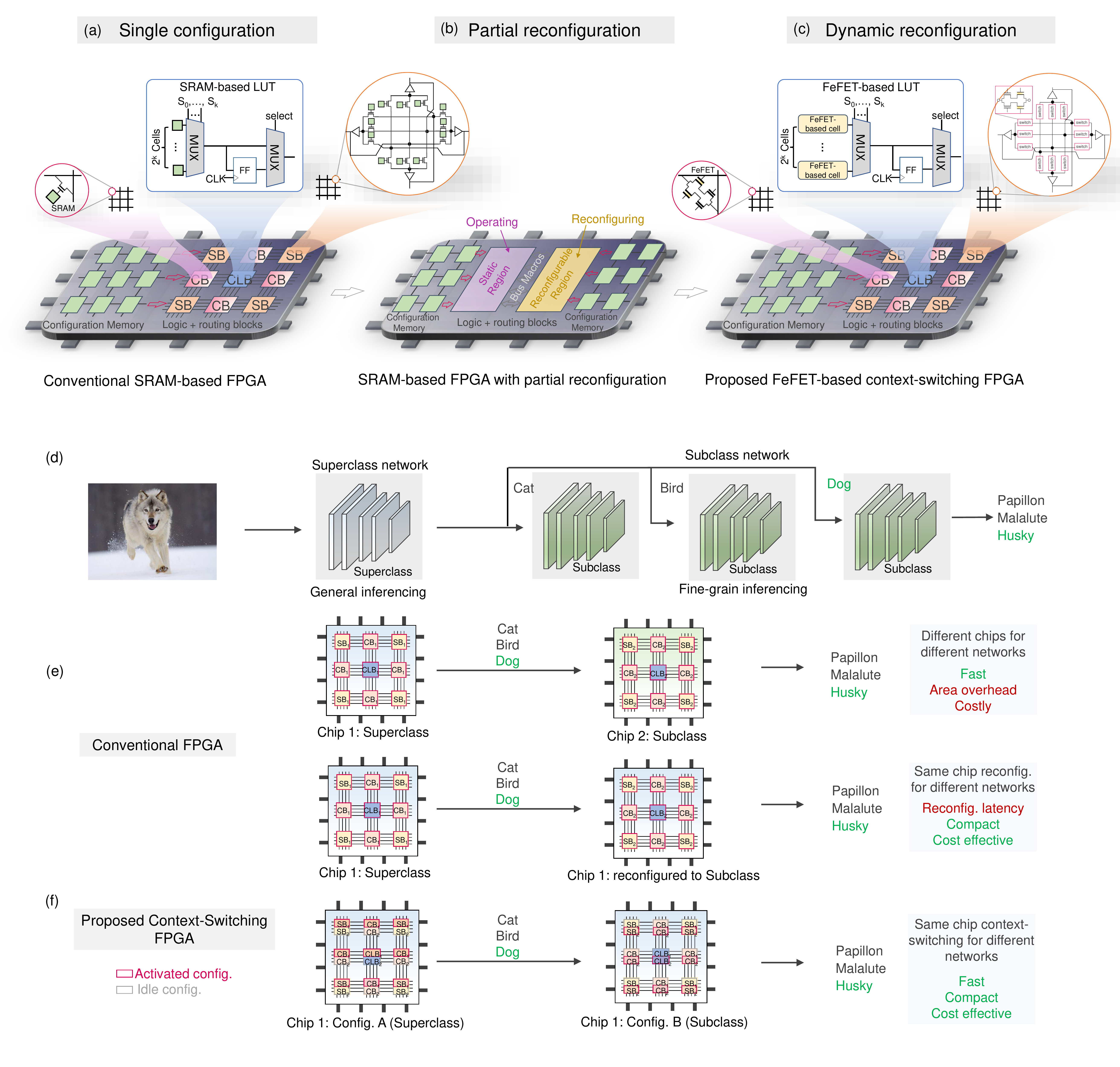}
    \captionsetup{parbox=none} 
    \caption{\textbf{Overview of the proposed context-switching FPGA and potential applications.} Architectures of (a) a conventional SRAM-based FPGA, (b) SRAM-based FPGA supporting partial reconfiguration, and (c) Our proposed FeFET-based context switching FPGA supporting dynamic reconfiguration. (d) An example of deep learning network: Two-stage Super-Sub network\cite{Shixian'USC}, where at first the superclass 'Dog' is identified and then the subclass 'Husky' is identified. (e) Conventional FPGA incurs area overhead or significant reconfiguration latency. This figure shows two main approaches of implementing the Super-Sub network in conventional FPGA. (f) Our approach provides fast reconfiguration speed and compact solutions.}
    \label{fig:DLN}
\end{figurehere}

As a concrete and highly important example of DNN acceleration on FPGA, a two-stage Super-Sub network is adopted for image classification\cite{Shixian'USC}. In this model, a superclass is first inferred using a generalist superclass-level network and the network output is then passed to a specialized network for final subclass-level classification. In this way, the overall classification accuracy has been proved to increase over that of common inference methods when evaluating on the "Superclassing ImageNet dataset", which is a subset of ImageNet\cite{ILSVRC15} and consists of 10 superclasses, each containing 7-116 related  subclasses  (e.g., 52 bird types, 116 dog types)\cite{Shixian'USC}. Fig.~\ref{fig:DLN}(d) shows one specific example of this framework. In the first stage, the superclass 'Dog' is identified by the generalist superclass network. Then, fine-grain inference in the subclass network is performed in the second stage and outputs the final result 'Husky' of the target image.

Numerous hardware accelerators have been proposed to implement DNNs, such as customized application-specific integrated circuits (ASICs)\cite{ASIC'17, ASIC'18, ASIC'22}, application-driven optimization on graphics processing units (GPUs)\cite{nvidia,nvidia_T4}, and FPGAs\cite{FPGA'16, FPGA'17}. 
However, among these various types of DNN accelerators, FPGA, which can provide more flexibility while maintaining high performance, is particularly suitable for implementing the accelerators of DNNs such as for the Super-Sub network model. Fig.~\ref{fig:DLN}(e) shows two main approaches when considering to implement this Super-Sub network into FPGA. One distinguished feature of the implementation is the requirement of multiple configurations in FPGA to map the superclass and sub networks, respectively. 
The straightforward approach is to use more than one chips to process different networks (i.e., configurations). As shown in Fig.~\ref{fig:DLN}(e), Chip 1 is configured to process the general inference task for superclasses, whose outputs are then sent to the Chip 2 which is configured to map the subclass networks to identify the specific subclass. This approach, although fast, incurs penalties in chip area and cost. Another compact and cost-efficient approach is to leverage the reconfiguration capability of FPGA by simply reconfiguring Chip 1 to the subclass network after it finishes execution of the superclass network. 
In this way, contexts, i.e., FPGA configurations, can be swapped in or out of the FPGA upon the demands of application requirements without the need of additional chips\cite{fpga1998}. Therefore, this approach saves the area cost but comes with a penalty in the reconfiguration latency. Above all, although FPGA offers an attractive choice for acceleration of  Super-Sub network model (Fig. ~\ref{fig:DLN}(e)), an ideal implementation with  high area efficiency and low latency is still elusive with current FPGA technologies and architectures.

Many relevant works have explored design options to address the aformentioned issues at different granularity of reconfiguration and from different angles of applications. However, all of them are still limited by the dilemma or might incur other overheads.
For example, a full context-switching FPGA was first proposed as a time-multiplexer FPGA based on the Xilinx XC4000E FPGA in 1997\cite{FPGA'1997}, where eight configurations of the FPGA are stored in on-chip memory and the contexts can be switched in a single cycle. With pre-loaded contexts, reconfiguration is not needed but it comes with a large area penalty. The more configurations to be supported, the more area overhead to store those configurations. In order to save area while still speeding up the reconfiguration process, dynamic partial reconfiguration appears as another solution to support multiple configurations, by which only a portion of hardware region (called reconfigurable region) can be reconfigured while the remainder is static\cite{CCS'18}. 
Partial reconfiguration brings several advantages over conventional context-switching FPGA\cite{Babu2021}, including less reconfiguration time compared to full-region reconfiguration and smaller area with its increased logic density. 
However, partial reconfiguration only provides a compromised solution between the area cost and the reconfiguration latency, incapable of fundamentally solving the problem.
At the end, it is possible to support fine-grain reconfiguration at bit level, 
as demonstrated by consecutive works on the 'NATURE' FPGA architecture to support fine-grain temporal logic folding\cite{nature_nano,sram'2011}, which is either based on CMOS (e.g., logic and SRAM) and carbon nanotube random-access memory (NRAM)\cite{nature_nano}, or based entirely on CMOS circuits \cite{sram'2011}. In the former work, NRAM and SRAM work together to support dynamic reconfiguration for temporal logic folding of circuits, which is to realize different logic functions in the same logic elements through dynamic reconfiguration every few cycles, thereby significantly increasing the logic density.
In the latter work, the dynamic reconfiguration delay is hidden behind the computation delay through the use of shadow SRAM cells (i.e., two SRAM copies). However, both works suffer from high area cost which is mainly caused by extra NRAM cells and 10T-SRAM cells respectively. Therefore, to date, a context-switching FPGA that can break the trade-off between the area cost and the reconfiguration latency remains elusive and the goal of the proposed research is to bridge the gap.

To mitigate the aforementioned issues in terms of area, latency and power, we propose a novel dynamic context-switching FPGA architecture based on FeFETs which can implement DNN accelerators more efficiently. With joint innovations from technology, circuit, and architecture levels, our proposed design has several advantages over prior context-switching works. First, from technology's perspective, FeFET is unique that it behaves both as a transistor switch and a nonvolatile memory cell such that FPGA basic logic circuits (e.g., LUTs) and routing elements (e.g., CBs and SBs) can be implemented compactly. Moreover, these FPGA basic elements have no leakage power dissipation because of the non-volatility of FeFET, which hugely reduces the total power consumption of the entire FPGA.
Second, from circuit's perspective, a new CB composed of two parallel branches is proposed, which stores two configurations while still consuming much less area than a single configuration SRAM-based CB. Third, the proposed FPGA is dynamically reconfigurable with the capability to load one configuration  without interrupting execution of another configuration. As a result, the reconfiguration time can be completely hidden as long as it is smaller than the computation time of the current active configuration. Therefore, our proposed solution can achieve dynamic context-switching with zero penalty in reconfiguration latency and significant area reduction compared to SRAM-based design, breaking the trade-off between area cost and reconfiguration latency existed in conventional CMOS implementations. 

With the proposed context-switching FPGA, the aforementioned Super-Sub network can be efficiently implemented, as shown in Fig.~\ref{fig:DLN}(f).
Considering one case that we are interested in having an accurate classification of one specific superclass (e.g., Dog), the proposed design can perfectly fit in it and reduce the reconfiguration latency. Specifically, these two configurations including superclass network and subclass network can be preloaded into the FPGA. First, the general inference with the superclass network is performed. As long as the output of the general inference is Dog, the configuration corresponding to Dog's subclass network would be activated and executed for further inference. In this way, compared to long reconfiguration time, the switching time is much less or even negligible, which leads to almost zero latency overhead. In addition, the total area cost could also be heavily reduced by leveraging dense FeFETs.
Note that the proposed context-switching FPGA enables applications in various domains that need switching between different contexts, beyond the Super-Sub network discussed here. The reconfiguration functionality is especially helpful in various dynamic adaptation applications such as changing communication encoders or decoders on demand to the appropriate protocols\cite{app_encoder_decoder}, changing the data rates to vary bandwidths\cite{app_datarate}, scaling the computation based on available energy needs\cite{app_energy}.
Moreover, with no limitation of the number of configurations, our design can also be scaled to implement multiple configurations depending on the demand of applications. Some potential applications are illustrated in Supplementary Fig.~\ref{fig:supplementary_app}.

\section*{\large Overview of the proposed FPGA architecture}

For a deeper look into the design of the proposed context-switching FPGA, details of the architecture and components to support multiple configurations are shown in Fig.~\ref{fig:FPGA_system}. Fig.~\ref{fig:FPGA_system}(a) shows primitive components of the proposed context-switching FPGA which supports dual configurations, including CLBs, CBs and SBs. 
For each component, it is controlled by the configuration information stored in configuration memory. By loading the configuration bits, the logic (LUT) and routing elements (CB/SB) can be connected  to form a functional circuit to  perform the desired computation.
In the proposed context-switching FPGA, there are two local copies of each LUT, CB and SB, which corresponds to two configurations. In this way, when one configuration is active for computation, any other configuration can be loaded without interrupting the execution, thereby significantly reducing the reconfiguration latency.
In contrast, in conventional context-switching FPGA, they would either require hardware resources for supporting multiple configurations on-chip or require long serial reconfiguration time. 
To support run-time reconfiguration and reduce the area cost incurred by the need of an extra copy of FPGA primitive components, FeFET technology, due to its programmability, nonvolatility, and compactness, is chosen in this work to implement basic programmable FPGA components such as LUTs, CBs and SBs.

\begin{figurehere}
   \centering
    \includegraphics[scale=1,width=\textwidth]{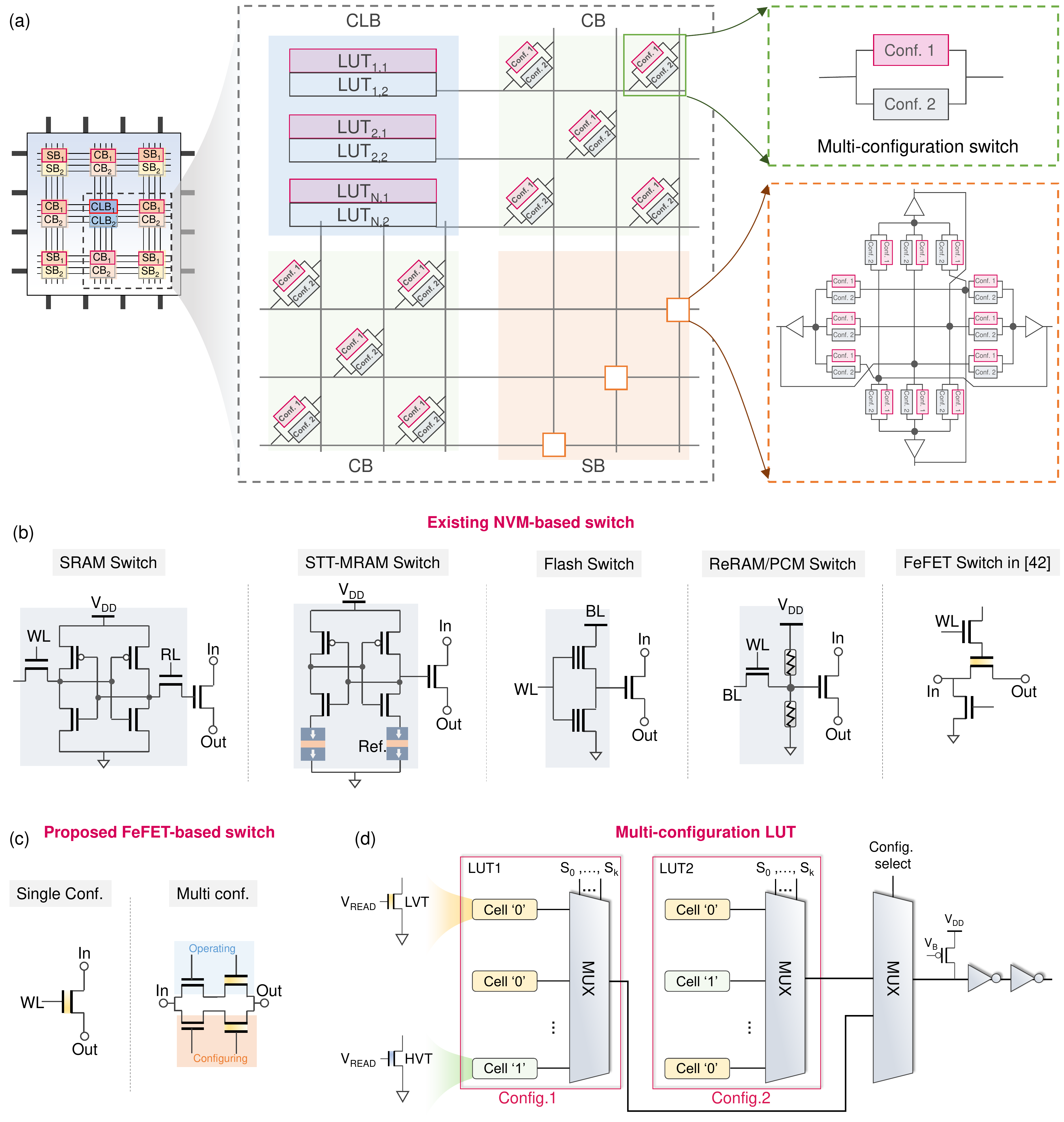}
    \captionsetup{parbox=none} 
    \caption{\textbf{The proposed dynamic context-switching FPGA architecture.} (a) Primitive FPGA components with dual configuration support. (b) Existing memory technology-based single configuration switch implementations. (c) Proposed FeFET-based switches. In the multi-configuration switch, we achieve dynamic reconfiguration by turning the pass transistors on/off to select active branch/reconfigure branch. (d) Proposed FeFET-based LUT for dual-configuration. Basically, it consists of two single configuration LUTs and one extra multiplexer for selecting proper configuration when needed.}
    \label{fig:FPGA_system}
\end{figurehere}

In recent years, the switches in FPGA can be realized with various embedded memory technologies as the basic elements of routing elements (CBs and SBs). Fig.~\ref{fig:FPGA_system}(b) presents existing mainstream memory technology-based single configuration switches including SRAM, spin transfer torque magnetic RAM (STT-MRAM), Flash memory, resistive RAM (ReRAM), phase change memory (PCM) and FeFET. Due to its logic-compatibility, superior write and read performance, and excellent reliability, SRAM is the most straightforward memory to use by combining a SRAM cell with an \textit{N}-type pass transistor. However, SRAM-based switches suffer from two crucial overheads. One is low area density due to its complex cell structure; the other is high leakage power, which accounts for 60\% $\sim$ 70\% of total FPGA power dissipation due to long routing tracks\cite{10.1145/968280.968285, 10.1145/503048.503072, 10.1145/611817.611844,10.5555/647929}.
Recently, emerging embedded nonvolatile memory technologies have been actively investigated as promising alternatives to SRAM due to their  density, energy, and performance advantages. However, each of them comes with its own challenges. For example, a Flash memory-based switch is nonvolatile and compact\cite{greene201165nm}, but memory programming is slow ($\sim$ms) and requires a high programming voltage ($\sim$10 volts). Two terminal resistive memories, including ReRAM, PCM, and STT-MRAM, are nonvolatile and dense, but usually require a large conduction current to program the devices, consuming a significant write power. 
Additionally, the limited on/off resistance ratio ($\sim$100 for ReRAM/PCM and $\sim$5 for STT-MRAM) usually requires additional circuitry, such as the 1T2R structure for ReRAM/PCM \cite{ tanachutiwat2010fpga, huang2014low} and an even more complex supporting structure for STT-MRAM \cite{zhao2009spin} to realize a single switch. 

In this regard, we propose the FPGA architecture which adopts FeFETs to implement logic and routing elements. Ever since the discovery of ferroelectricity in doped HfO\textsubscript{2}, significant progress has been made in the integration of HfO\textsubscript{2} based FeFET due to its nonvolatility, high density, large ON/OFF ratio, and excellent CMOS-compatibility\cite{mulaosmanovic2021ferroelectric, schroeder2022fundamentals}. In addition, switching of ferroelectric polarization is induced by an applied electric field, rather than a large conduction current, making FeFET a highly energy-efficient nonvolatile memory\cite{khan2020future}. Since the ferroelectric film is integrated in the gate stack of a FeFET, when its polarization is set to point at the channel/metal gate, the FeFET threshold voltage (\textit{V}\textsubscript{TH}) will be programmed to the low-\textit{V}\textsubscript{TH}/high-\textit{V}\textsubscript{TH}, respectively, thus realizing a compact nonvolatile routing element\cite{yu2022hardware}. Leveraging this technology, a mixed FeFET/CMOS switch unit (i.e., 2T-1FeFET) has been proposed as a routing element in FPGA\cite{TCSI'2018}, which takes advantage of but does not fully exploit FeFET. In this work,
leveraging the intrinsic nonvolatile switch structure of FeFET, we propose a 1FeFET routing switch for single configuration FPGA and a 2T-2FeFET routing switch for dynamic reconfiguration context-switching FPGA, as shown in Fig.~\ref{fig:FPGA_system}(c), which achieve optimal area efficiency. For the context-switching FPGA, a serial CMOS transistor is added to each branch, which is used to cutoff the branch that is loading a new configuration to minimize the disturb to the other active branch. Fig.~\ref{fig:FPGA_system}(d) shows our proposed circuit of LUT array for dual configuration. A compact LUT cell can be efficiently implemented using a single FeFET such that the high-\textit{V}\textsubscript{TH}/low-\textit{V}\textsubscript{TH} states of FeFET stores bit '1'/'0' for the LUT cell, respectively. 
Besides, as shown in Fig.~\ref{fig:FPGA_system}(d), the proposed LUT can support dynamic reconfiguration -- when the branch of configuration 1 is operating, the branch of configuration 2 can load new configuration.

\section*{\large Block Design and Functional Verification}

In this section, experimental verification of the proposed LUT and routing elements (CB/SB) for context-switching FPGA is performed. For experimental demonstration, FeFET devices integrated on the 28 nm high-$\kappa$ metal gate (HKMG) technology are tested. Fig.~\ref{fig:LUT}(a) and (b) show the transmission electron microscopy (TEM) and schematic cross-section of the device, respectively. The device features an 8 nm thick doped HfO\textsubscript{2} as the ferroelectric layer and around 1 nm SiO\textsubscript{2} as the interlayer in the gate stack. The FeFET memory performance is characterized by standard pulsed \textit{I}\textsubscript{D}-\textit{V}\textsubscript{G} measurements after applying $\pm$4 V, 1$\mu$s write pulses on the gate. Fig.~\ref{fig:LUT}(c) shows a memory window about 1.2 V, i.e., the \textit{V}\textsubscript{TH} separation between the low-\textit{V}\textsubscript{TH} and high-\textit{V}\textsubscript{TH}
states, which enables a large ON/OFF conductance ratio. It also exhibits a well-tempered cycle-to-cycle variation. Fig.~\ref{fig:LUT}(d) shows the switching dynamics of the FeFET under different pulse amplitudes and pulse widths, which also shows a trade-off between the write speed and pulse amplitude and that it is possible to program FeFET with sub-10ns with 4V write amplitude. It follows the classic nucleation-limited switching model in the thin film poly-crystalline HfO\textsubscript{2} \cite{mulaosmanovic2017switching,mulaosmanovic2021ferroelectric}, where domain switching is mainly limited by the nucleation process and the nucleation time follows an exponential dependence on the applied electric field. These results suggest that HfO\textsubscript{2} based FeFET exhibits a high performance, showing great promise of this technology in many applications including the context-switching FPGA in this work.  

Fig.~\ref{fig:LUT}(e) and (f) show the operation principle of our proposed LUT cells that store a bit '1' and '0', respectively. Each cell consists of one single FeFET and one PMOS transistor, where the PMOS is shared among all the cells and is part of the sense amplifier used to convert the read current to logic voltage levels. The bit '1' and '0' is stored by programming the FeFET into the high-\textit{V}\textsubscript{TH} and low-\textit{V}\textsubscript{TH} state, respectively. 
Then in the LUT read mode, the stored bit can be read by asserting appropriate read voltage, \textit{V}\textsubscript{READ}, to the gate terminal of the FeFET, as shown in Fig.~\ref{fig:LUT}(e). Due to the large ON/OFF resistance ratio of FeFET at \textit{V}\textsubscript{READ}, the output voltage will be close to \textit{V}\textsubscript{DD} and ground for bit '1' and '0', respectively. This is achieved
by choosing an appropriate PMOS gate bias (\textit{V}\textsubscript{B}) such that its resistance is between the FeFET high-\textit{V}\textsubscript{TH} and low-\textit{V}\textsubscript{TH} states, thereby setting the output voltage rail-to-rail.  
Fig.~\ref{fig:LUT}(g) demonstrates the main structure of the single configuration LUT integrated with $2^{k}$ FeFET-based bitcells (Cell '0'/Cell '1'), different logic functions can be successfully achieved by applying different combinations of select signals. In this structure, a sense amplifier composed of one pull-up PMOS transistor and two inverters is used for converting FeFET read current to voltage and amplifying the output voltage to full swing. The LUT cell operation is then verified in experiment using the setup shown in Fig.~\ref{fig:LUT}(h), which includes the major components in Fig.~\ref{fig:LUT}(g). The operation waveforms are presented in Fig.~\ref{fig:LUT}(i), which shows the write and read phases of the LUT cell. After programming the FeFET into high-\textit{V}\textsubscript{TH}/ low-\textit{V}\textsubscript{TH} states using -4 V/+4 V, 1$\mu$s write pulse, the output voltage shows a logic high and low, respectively. This verifies the successful cell operation, but due to the discrete experimental setup, performance is limited by the parasitics. In order to predict the fully-integrated FeFET LUT performance, SPICE simulations using a calibrated FeFET model\cite{fefet} and 45 nm Predictive Technology Model for logic transistor (PTM\cite{ptm}) are performed. Supplementary Fig.~\ref{fig:supplementary_lut} shows the simulated waveform, indicating that for a 6-input LUT cell, the read delay is 124.3ps and consumes 13.1 $\mu$W power. In the subsequent section, FeFET based primitive components, including LUTs, CBs, and SBs, are also compared with other technology implementations using consistent SPICE simulations, as will be studied in Fig.\ref{fig:table}.

\begin{figurehere}
   \centering
    \includegraphics[scale=1,width=0.9\textwidth]{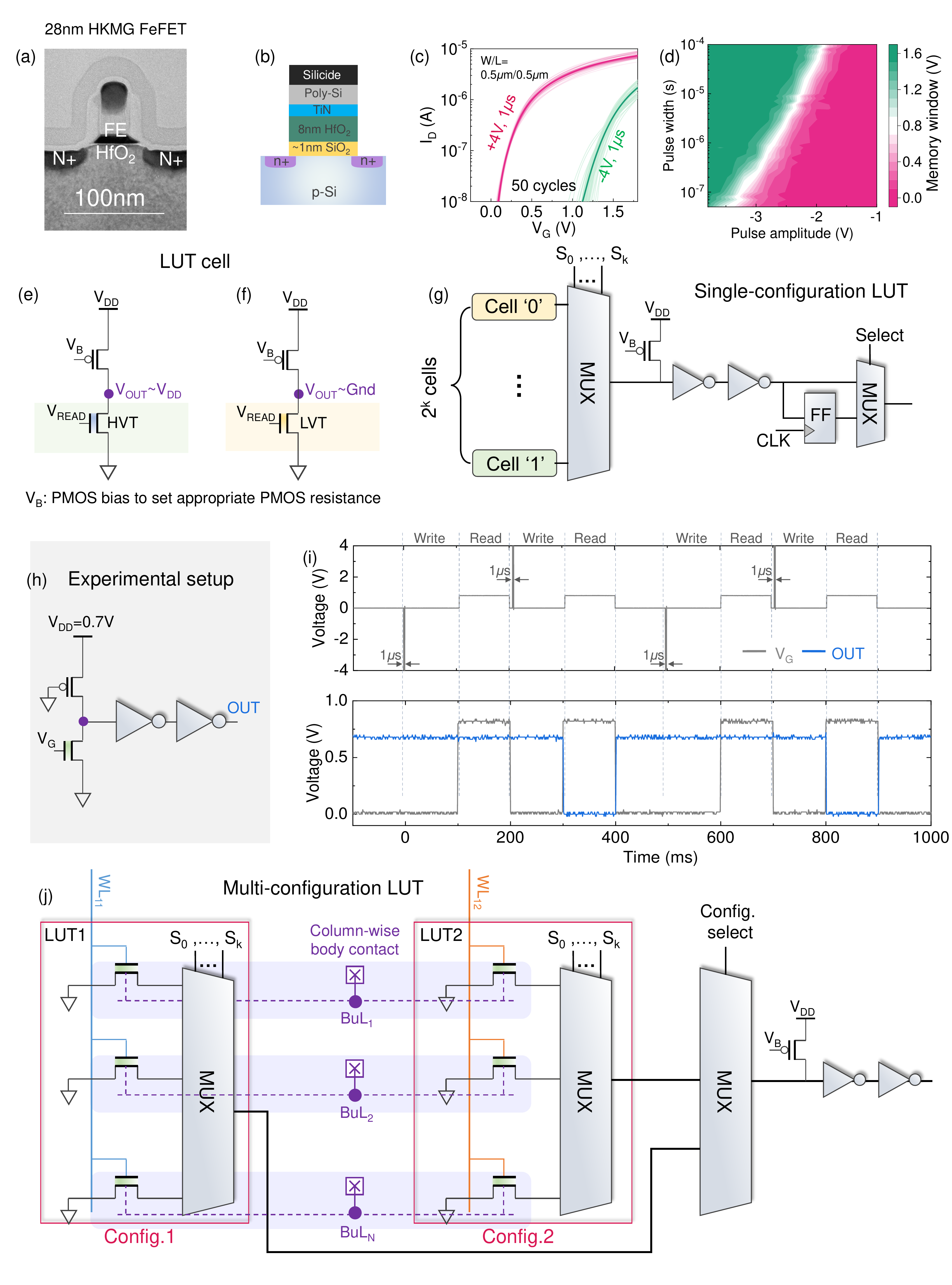}
    \captionsetup{parbox=none} 
    \caption{\textbf{Experimental verification of the LUT cell operation.} (a-b) TEM and schematic cross section. (c-d) $I_{D} –V_{G}$ characteristics for FeFET measured after ±4 V, 1 $\mu$s write pulses and the switching dynamics of FeFET under different pulse amplitudes and pulse widths. (e-f) Operations of the LUT cell for storage with bit ‘0’/’1’ by exploiting the dynamic LVT/HVT programming capability. (g) Proposed k-bit LUT. (h) The experimental setup of functional verification of the LUT cell operation. (i) Experimental waveforms of proposed LUT cells in (e-f). (j) The circuitry of a LUT array for multiple configuraitons.}
    \label{fig:LUT}
\end{figurehere}

To support dynamic reconfiguration, two LUTs forming an array are designed and an additional multiplexer is used to select which configuration should be active in current operating period, as shown in Fig.~\ref{fig:LUT}(j). Programming in a bulk planar single FeFET array has been extensively investigated\cite{jiang2022on, xiao2022write}. The applicable programming schemes depend on the number of accessible terminals during memory write. In the proposed FPGA architecture, the source/drain terminals are not simultaneously accessible from outside, which limits the possibility of applying write schemes that need to apply the source/drain voltages. In this case, a convenient solution is shown in Fig.~\ref{fig:LUT}(j), where the gate and the body terminals are used for programming. The word line (WL) is shared among all FeFETs in a configuration block and the body is shared across different configuration blocks. Two step programming will then be performed where all the FeFETs in a configuration are set to the low-\textit{V}\textsubscript{TH} states first by applying a positive write voltage (i.e., \textit{V}\textsubscript{W}) on the WL and keep all the other terminals grounded. Then those FeFETs need to be in the high-\textit{V}\textsubscript{TH} states are applied with a negative gate-to-body voltage (i.e., -\textit{V}\textsubscript{W}). To avoid write disturb to those low-\textit{V}\textsubscript{TH} states FeFETs during the second step, the standard inhibition bias scheme (e.g., \textit{V}\textsubscript{W}/2) can be applied, which is verified in Supplementary Fig.\ref{fig:supplementary_write_disturb}

\begin{figurehere}
   \centering
    \includegraphics[scale=1, width=1\textwidth]{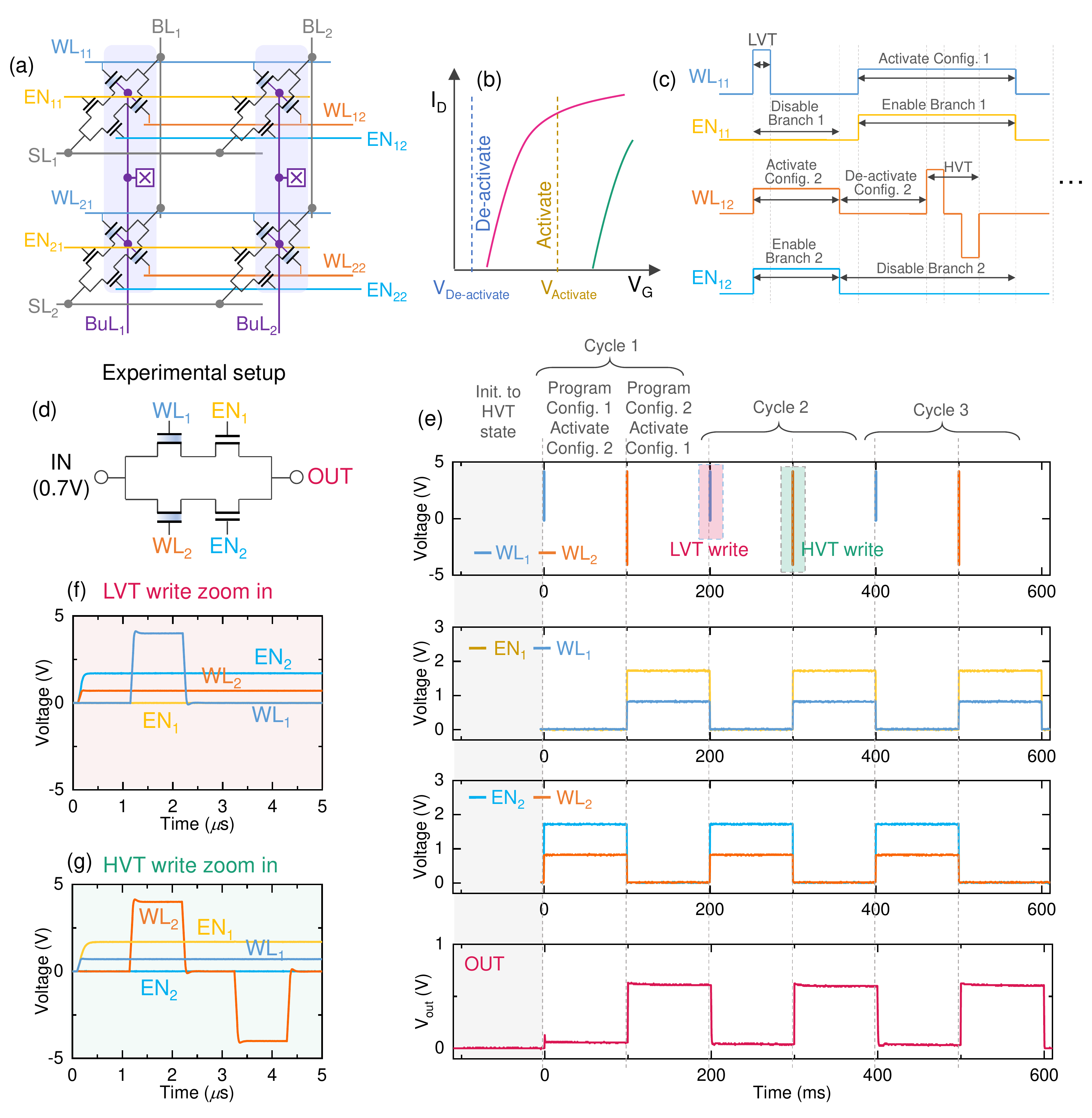}
    \captionsetup{parbox=none} 
    \caption{\textbf{Experimental verification of the multi-configuration CB operation.} (a) The structure of one 2x2 CB array. (b) By applying different read gate voltages, the swap between configurations can be achieved. (c) An example waveform applied to set the branch 1/branch 2 to be at the low-\textit{V}\textsubscript{TH}/high-\textit{V}\textsubscript{TH} states respectively without interrupting normal operation. (d) The circuitry of one CB test unit. (e) The experimental transient waveforms of run-time context configuration and switching repeated for 3 cycles. (f)/(g) The zoomed-in programming waveform for branch 1/branch 2 in tests, respectively. The zoomed-in programming waveform is shown due to its small write pulse width.}
    \label{fig:CB}
\end{figurehere}

Next the functionality of the routing elements is verified, as shown in Fig.\ref{fig:CB}. Using CB as an example, Fig.\ref{fig:CB}(a) shows the array structure, where bit line (BL) and source line (SL) route the actual signal, and WL and the column-wise body contact are used to program FeFETs. As introduced in Fig.\ref{fig:FPGA_system}(c), to support the run-time reconfiguration of one branch without interrupting the normal operation of the other branch, a serial transistor is added to each branch and is off/on during configuration loading/execution, respectively. The swap between configurations can be easily and swiftly conducted by applying corresponding read gate biases, as shown in Fig.\ref{fig:CB}(b), such that when one configuration is de-activated, the FeFET will be cut-off, irrespective of its states. 
Fig.\ref{fig:CB}(c) shows an example waveform applied on a testing unit (Fig.\ref{fig:CB}(d)), where the branch 1 is first configured to be the low-\textit{V}\textsubscript{TH} state while branch 2 is executed and then branch 1 is activated while the branch 2 is configured to the high-\textit{V}\textsubscript{TH} state using the two-step programming. Fig.\ref{fig:CB}(e) shows the experimental results applied the voltage sequence shown in Fig.\ref{fig:CB}(c) for three repeated cycles. The zoomed-in programming waveforms for branch 1 and branch 2 are shown in Fig.\ref{fig:FPGA_system}(f) and (g), respectively. Due to the configurations used in this testing scenario, where the branch 1/branch 2 is in the low-\textit{V}\textsubscript{TH}/high-\textit{V}\textsubscript{TH} states respectively, the output signal will therefore switch between 0.7 V (i.e., when branch 1 is active) and 0 V (i.e., when branch 2 is active). The experimental results therefore confirm successful operations. Supplementary Fig.~\ref{fig:cb_hvt_hvt}-Fig.~\ref{fig:cb_hvt_lvt} show experimental results of the other three configuration combinations of two branches, which further verifies the successful run-time reconfiguration operation. Similar to the LUT cell case, SPICE simulations are conducted to predict the speed of a fully integrated CB, where the simulated transient waveform of proposed multi-configuration CB is shown in Supplementary Fig.~\ref{fig:supplementary_cb}.

\section*{\large Evaluation and Application Case Study}

To evaluate the feasibility and performance of the proposed FeFET-based context-switching FPGA architecture, simulations are performed and a comprehensive comparison with other relevant works based on different memory technologies is shown in terms of area, delay and power consumption. Moreover, at the system level, the capability of the proposed architecture to successfully achieve dynamic reconfiguration is demonstrated and the evaluation results show that the design presents a significant power reduction and area efficiency improvement with slightly increased critical path delay as the trade-off.
To estimate the area of FeFET-based CB and LUT cell and compare with other works, the layouts are drawn and the area is calculated using lambda design rules in Supplementary Fig,.~\ref{fig:supplementary_layout}. All relevant area numbers are shown in Fig.~\ref{fig:table}(a). 
Our layout analysis shows that the proposed CB and LUT cell are more compact compared to SRAM CBs and LUT cells. For example, the proposed FeFET-based single configuration CB and LUT cell, occupy  area that is only 8.5\% and 18.5\% of their respective SRAM-based counterparts while the prior FeFET-based CB and LUT cell\cite{TCSI'2018} require 36.4\% and 36.2\%  of that area, respectively. 
Even the proposed multi-configuration FeFET CB and LUT cell area  is only  28.9\% and 37.0\% of that of the SRAM-based single configuration design.
Therefore, the proposed design shows a significant area reduction compared to SRAM-based design and previous FeFET-based design\cite{TCSI'2018}.

Fig.~\ref{fig:table}(b) summarizes the basic structures of 6-input LUT/CB/SB based on existing memory technologies (SRAM, STT-MRAM, RRAM and FeFET), and compares their corresponding read delay and read power consumption. All circuits are simulated with HSPICE. The 45nm Predictive Technology Model\cite{ptm} is adopted for all MOSFETs in this work and a calibrated FeFET model\cite{fefet} is used for the proposed design. For resistive memories, the corresponding low resistance and high resitance levels are used for simulation\cite{rram, stt}.
According to the simulation results (Fig.~\ref{fig:table}(b)), we observe that for a 6-input LUT, the proposed single configuration LUT shows the smallest read power consumption, which is 13.1 $\mu$W, and for multiple configurations, this number increases slightly but still less than the power consumed by MTJ-based single configuration LUT. This is due to the large on/off ratio of FeFET obviating the need for a high read current to differentiate its two states, unlike MTJ designs.
As for the read delay, RRAM-based single configuration LUT has the longest latency. The proposed FeFET-based single configuration LUT shows the second best latency in all considered nonvolatile LUTs. Besides, the delay of the proposed FeFET-based multi-configuration LUT is less than that of RRAM-based single configuration LUT even though considering one extra multiplexer for selecting configurations. The switching current through the sense amplifier for FeFET is larger than RRAM due to its higher on/off ratio (lower \textit{R}\textsubscript{on}), resulting in less LUT delay than RRAM. For CBs, our 1FeFET single configuration CB and 2T-2FeFET multi-configuration CB show much less power consumption during operation, which consume $\sim$95\%/$\sim$85\% less power than the SRAM-based CB. For SBs, both FeFET-based single configuration SB and multi-configuration SB show much less power consumption than others since our circuit contains less transistors. However, the delay of 1FeFET CB is around 2x times of that of a SRAM-based CB. The delay of FeFET-based SB is worst among different memory technology based designs. That is becuause FeFET's transmission speed is not so high as a conventional MOSFET, resulting in poorer performance as CB. In conclusion, the proposed FeFET-based designs (CB/SB) show significant advantages on power consumption over SRAM/STT-MRAM/RRAM based designs but with the slight penalty in delay. Note that the penalty in the routing elements' (CB/SB) delay does not necessarily mean that the overall system will be impacted as the routing delay may be a small portion of the overall system delay, which is investigated below (Fig.~\ref{fig:table}(c)).

\begin{figurehere}
   \centering
    \includegraphics[scale=1,width=\textwidth]{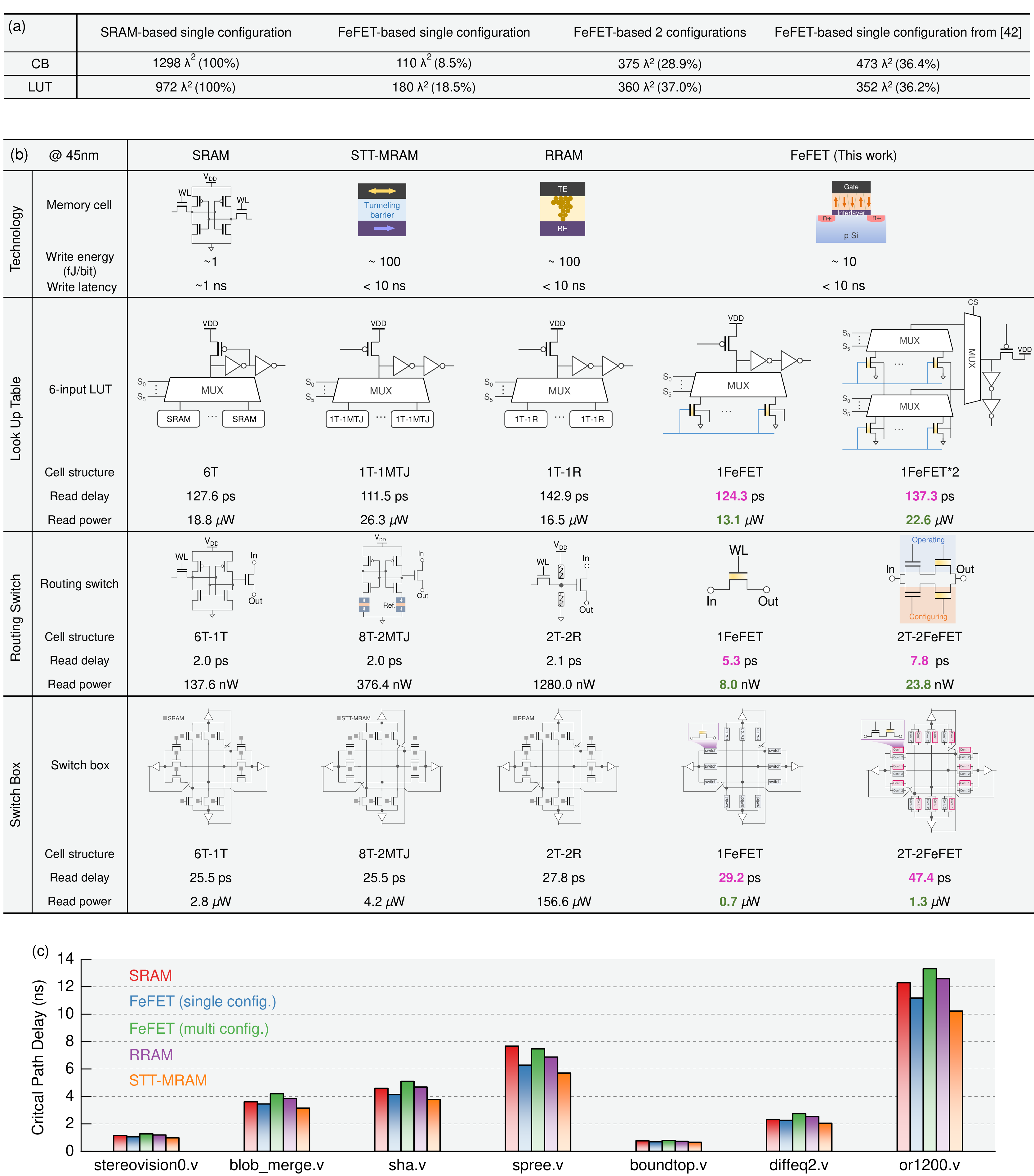}
    \captionsetup{parbox=none} 
    \caption{\textbf{Area comparison and simulation results.} (a) Area impact of FeFET LUT cells (storage) and CBs over SRAM based structures. (b) Delay and Power comparison of main components of FPGA based on different memory technologies. (c) Critical path delay of different memory technology-based FPGA designs.}
    \label{fig:table}
\end{figurehere}

In order to investigate the impact of the primitive (i.e., LUT/SB/CB) delay on the latency of the whole FPGA, the critical path delay is studied with the verilog-to-routing (VTR) tool\cite{vtr}. The VTR tool is a popular open source CAD tool for FPGA architecture development and evaluation. 
For fair comparison, all the SRAM-/RRAM-/STT-MRAM-/FeFET-based FPGAs employ a well-optimized and commercial FPGA architecture using 45nm technology in VTR. To get the critical path delay of different memory technology based FPGAs, 7 circuitry benchmarks (stereovision0, blob\_merger, sha, spree, boundtop, diffeq2, and or1200) included in VTR are conducted\cite{vtr, vtr_benchmark1}. These represent popular applications in diverse domains, such as image processing, math, cryptography and computer vision.
Fig.~\ref{fig:table}(c) compares the critical path delay measured from SRAM-/RRAM-/STT-MRAM-/FeFET-based FPGAs.
Compared with SRAM-based FPGA, the FeFET-based single configuration FPGA presents 8.6\% reduction in the critical path delay on average, and it is also better than RRAM-based architecture. However, the proposed FeFET-based multi-configuration FPGA shows 9.6\% increment in the critical path delay compared to SRAM-based FPGA.
The simulation confirms that the delay of LUTs is dominant in the overall delay of the entire FPGA, therefore explaining the aforementioned performance of these FPGAs.

In addition, to show the feasibility of implementing the whole design in deep learning applications, three case studies under different scenarios are investigated. 
The first case is presented to show the benefit provided by dynamic reconfiguration in image classification. In the evaluation, two approaches of inference are considered -- static inference and dynamic inference. For static inference, the input image is classified by the generalist classifier. However, for dynamic inference, the input image is first\ classified by the superclass classifier to identify the superclass. If the superclass is supported by the specialist subclass classifier network, then the configuration of the subclass classifier would be switched and executed for enhanced accuracy. Otherwise, a generalist classifier is invoked to complete the subclass identification. The whole workflow is shown in Fig.~\ref{fig:case_study}(a). Fig.~\ref{fig:case_study}(b) shows that dynamic inference for super class classification improves the accuracy by up to 3.0\% over static inference.
Only context-switching FPGA can efficiently realize dynamic inference.
In last two cases, the feasibility and advantages of the proposed design over the conventional FPGA design are evaluated in terms of timing when considering various application scenarios. Basically, three neural networks (ResNet50, CNV, and MobileNetv1) are deployed into FPGA through Xilinx Vitis AI platform\cite{vitisai}. 
In the second case study, a case scenario that needs to switch between two neural networks frequently (Fig.~\ref{fig:case_study}(c)) is considered. In conventional FPGA, it is necessary to load new configurations before switching  contexts, which is time consuming.
However, for this context-switching design, our approach can preload two configurations, and then freely switch between them without the reconfiguration latency. The switch time of the proposed design is less than 1 ns which is much smaller than reconfiguration time and the proposed design shows significant speed up (from 39.0\% to 97.5\% (Fig.~\ref{fig:case_study}(d))).
The last case study is related to dynamic reconfiguration. It is assumed that there are three different neural networks to implement and switch between. Thus, in this case, there would be six situations corresponding to six combinations of these three networks (ResNet50$\rightarrow$CNV$\rightarrow$MobileNetv1, ResNet50$\rightarrow$MobileNetv1$\rightarrow$CNV, CNV$\rightarrow$ResNet50$\rightarrow$MobileNetv1, CNV$\rightarrow$MobileNetv1$\rightarrow$ResNet50, MobileNetv1$\rightarrow$R\-e\-s\-N\-e\-t\-5\-0$\rightarrow$CNV, and MobileNetv1$\rightarrow$CNV$\rightarrow$ResNet50).
As is well-known, latency is one of the most critical criteria when evaluating a neural network accelerator. Hence, for all these six situations, the total consumed time, including both the execution time and the reconfiguration time for each network, is compared under two different conditions -- one is in conventional FPGA, the other is in the proposed architecture with dynamic reconfiguration. 
As shown in Fig.~\ref{fig:case_study}(e), as the capability of dynamic reconfiguration means that the architecture is able to operate and reconfigure simultaneously, some parts of or even the complete reconfiguration time of the following network can be overlapped and hidden by the execution time of current network, which helps to reduce the total latency. As shown in Fig.~\ref{fig:case_study}(f), the results demonstrate that the proposed design with dynamic reconfiguration offers time saving for all these situations which varies from 2.4\% to 37.4\%. One thing should be noticed is that the maximum time saving of the ideal case would be 50\%, in which the execution time of the first network is equal to the configuration time of the second network. The maximum improvement of the proposed design (37.4\%) is very close to this number. Additionally, the proposed FPGA architecture is adaptive to implement more deep learning frameworks, and the relevant improvements and benefits are investigated in Supplementary Fig.~\ref{fig:supplementary-case_study}.
Above all, the case studies demonstrate that the proposed FeFET-based context-switching FPGA design shows the best adaptability in various types of deep learning applications.

\begin{figurehere}
   \centering
    \includegraphics[scale=1,width=\textwidth]{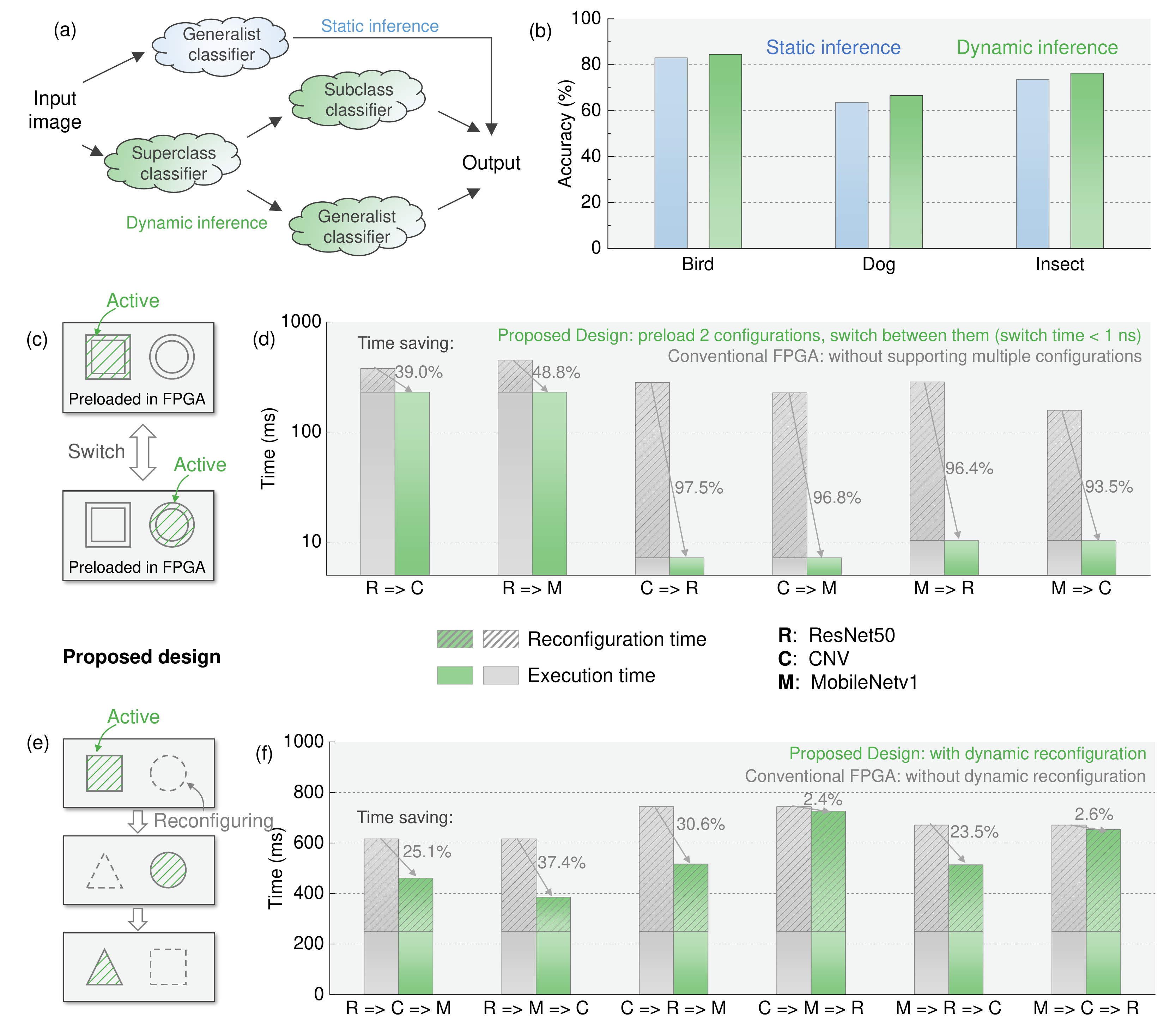}
    \captionsetup{parbox=none} 
    \caption{\textbf{Application case studies of the proposed multi-configuration FPGA for different application scenarios.} (a) Image classification workflow. (b) Dynamic inference for image classification improves the accuracy. (c) A diagram shows the experimental setup of the second case study: our design preloads two configurations in the FPGA, and then switch between them as needed. (d) Compared to conventional FPGA, the capability of switching between 2 configurations of our design yields significant time saving varying from 39.0\% to 97.5\% in our case (in an ideal case, the maximum time saving would be 100\%). (e) A diagram shows the experimental setup of the third case study: the proposed FPGA implements and performs three different neural networks using dynamic reconfiguration which achieves operating and reconfiguring simultaneously.  (f) Switching between 3 neural networks with dynamic reconfiguration offers time saving varying from 2.4\% to 37.4\% compared to traditional FPGA (in an ideal case, the time saving would be 50\%).}
    \label{fig:case_study}
\end{figurehere}
\section*{\large Conclusion}

In summary, we propose a novel FeFET-based context-switching FPGA architecture with the capability of dynamic reconfiguration, which can mitigate the tradeoff in conventional FPGA between the chip area cost and reconfiguration latency.    
In addition, we experimentally verify the functionality of the primitive blocks of the proposed FPGA. The simulation results reveal that by leveraging FeFETs, the proposed primitives of the FPGA show huge area and power reduction compared to conventional SRAM-based design. Moreover, three representative application scenarios are investigated and studied. The evaluation results show the proposed context-switching FPGA supporting dynamic reconfiguration offers significant time saving in these application scenarios. Our design provides an efficient solution to bridge the gap and makes FPGA more competitive in accelerating complex deep learning applications.

\section*{\large Data availability}
The data that support the plots within this paper and other findings of this study are available from the corresponding author on reasonable request.

\section*{\large References}
\bibliography{ref1.bib}

\section*{\large Acknowledgements}

This work is mainly supported by the U.S. Department of Energy, Office of Science, Office of Basic Energy Sciences Energy Frontier Research Centers program under
Award Number DESC0021118. The experimental characterization is partially supported by the Army Research Office under Grant Number W911NF-21-1-0341. The architecture benchmarking is partially supported by ND EPSCoR and partially by NSF 2132918 and 2008365. The device fabrication is funded by the German Bundesministerium für Wirtschaft (BMWI) and by the State of Saxony in the frame of the “Important Project of Common European Interest (IPCEI)".

\section*{\large Author contributions}

V.N., S.G. and K.N. proposed and supervised the project. Y.X., Y.X., T.Y., and E.Homan conducted circuit and architectural simulations. Z.Z. performed experimental verification. H.M., Y.S.K., S.D., and S.B. fabricated the FeFET devices. X.G. helped with FeFET array programming. R.J. and X.S.H helped with FPGA benchmarking. S.W., A.S.R., K.L., and L.I. conducted the benchmarking of FPGA implementation of super-sub networks. All authors contributed to write up of the manuscript.

\section*{\large Competing interests}
The authors declare no competing interests.

\newpage

\renewcommand{\thefigure}{S\arabic{figure}}
\renewcommand{\thetable}{S\arabic{table}}
\setcounter{figure}{0}
\setcounter{table}{0}

\newpage
\begin{center}
\title{\textbf{\Large Supplementary Materials}}
\end{center}

\section*{\large Device Fabrication}
In this paper, the fabricated ferroelectric field effect transistor (FeFET) features a poly-crystalline Si/TiN/doped HfO\textsubscript{2}/SiO\textsubscript{2}/p-Si gate stack. The devices were fabricated using a 28nm node gate-first high-$\kappa$ metal gate CMOS process on 300 mm silicon wafers. Detailed information can be found in \cite{trentzsch201628nm, fefet}. The ferroelectric gate stack process module starts with growth of a thin SiO\textsubscript{2} based interfacial layer, followed by the deposition of the doped HfO\textsubscript{2} film. 
A TiN metal gate electrode was deposited using physical vapor deposition (PVD), on top of which the poly-Si gate electrode is deposited. The source and drain n+ regions were obtained by phosphorous ion implantation, which were then activated by a rapid thermal annealing (RTA) at approximately 1000 $^\circ$C. This step also results in the formation of the ferroelectric orthorhombic phase within the doped HfO\textsubscript{2}.
For all the devices electrically characterized, they all have the same gate length and width dimensions of 0.5$\mu$m x 0.5$\mu$m, respectively.

\section*{\large Electrical Characterization}
The experimental verification was performed with a Keithley 4200-SCS Semiconductor Characterization System (Keithley system), a Tektronix TDS 2012B Two Channel Digital Storage Oscilloscope (oscilloscope), and a Keysight 81150A Pulse Function Arbitrary Generator (waveform generator). Two 4225-PMUs (pulse measurement units) were utilized to generate proper waveforms. The FeFETs used in experimental verification were connected with devices (inverters, p-type MOSFET, and/or n-type MOSFET) externally on a breadboard. In the experimental verification of the LUT cell operation, \textit{V}\textsubscript{DD} was given by the waveform generator. Output pulses were captured by the oscilloscope. Write and read operations were provided by the Keithley system. In the experimental verification of the multi-configuration CB operation, input voltage was given by the waveform generator. Output pulses were captured by the oscilloscope. WL and EN signals were generated by the Keithley system. Three repeated cycles were performed for each configuration. State initialization (+4V or -4V to both WL\textsubscript{1} and WL\textsubscript{2} with pulse width 1$\mu$s) was added at the beginning of the waveforms in order to generate a desired output in the first cycle. 

\newpage
\section*{\large Potential Applications}

\begin{figurehere}
   \centering
    \includegraphics[scale=1,width=1\textwidth]{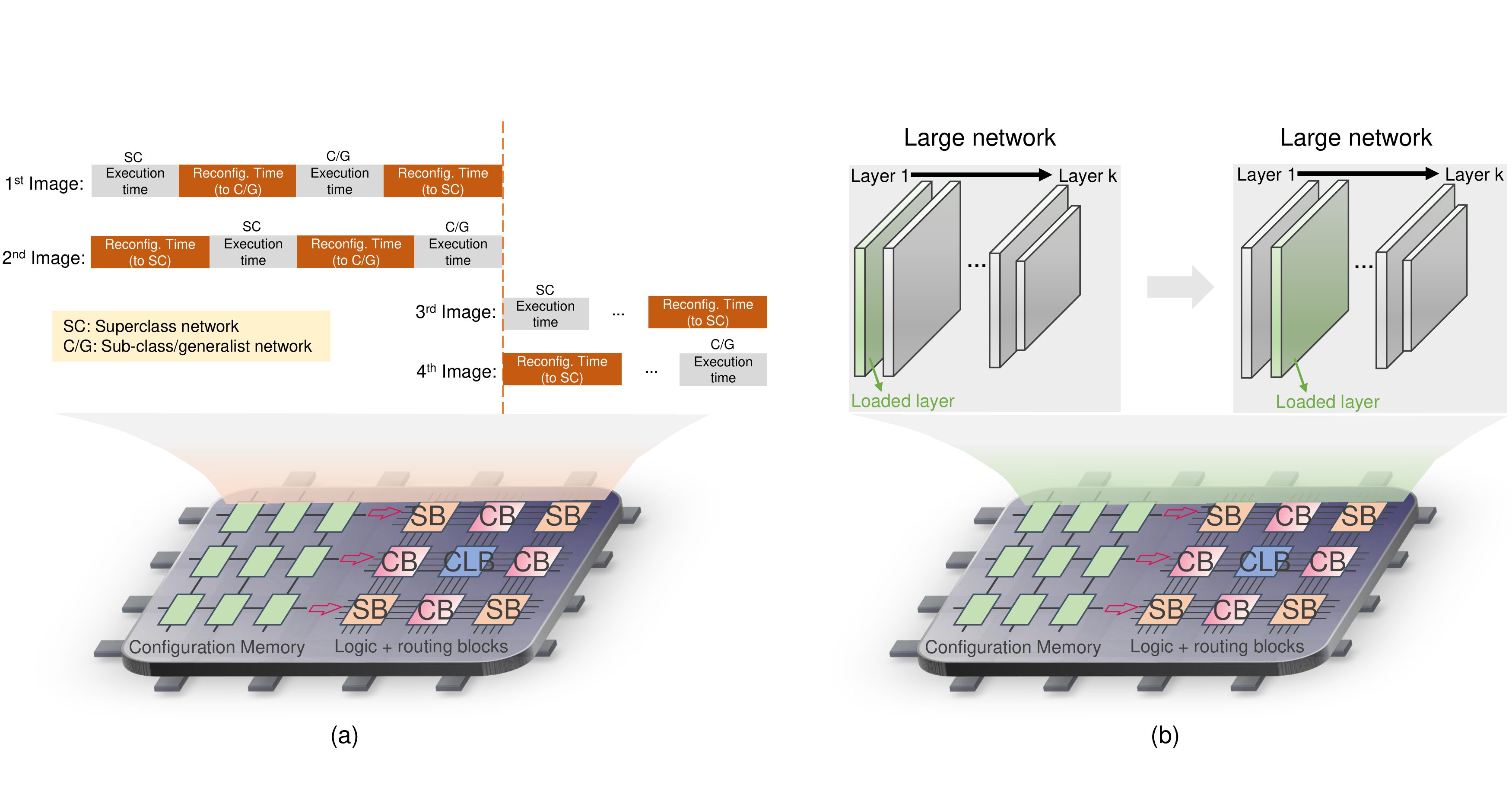}
    \captionsetup{parbox=none} 
    \caption{\textbf{Two potential applications of the proposed FeFET-based context-switching FPGA architecture.} (a) Our design can be used in image classification and help to reduces processing time dramatically for processing a large number of images.  (b) For some large and complex neural networks which can't completely fit in general FPGA, the proposed FeFET-based context-switching FPGA architecture provides reliable solutions through dynamic reconfiguration.}
    \label{fig:supplementary_app}
\end{figurehere}

In addition to the Super-Sub network application mentioned before, there are still a large number of deep learning applications which the proposed FeFET-based context-switching FPGA architecture can be suitable for or provide reliable solutions. In Fig.~\ref{fig:supplementary_app}, two potential application situations are presented. One is a derivative situation of the Super-Sub network application. When there are a large number of images needed to be classified, conventional FPGA without dynamic reconfiguration would inevitably require a extremely long time to process all these images due to the serial process mechanism. However, for the context-switching FPGA enabling dynamic reconfiguration, the processing time can be reduced dramatically since the proposed design supports multiple configurations and enables the capability of reconfiguring and executing simultaneously. More specifically, as shown in Fig.~\ref{fig:supplementary_app}(a), the proposed design only requires eight cycles to finish the task of image classification of four images while conventional FPGA would require more than sixteen cycles in the same situation.

The other potential application situation is for those large and complex neural network implementation. In recent years, with the increasing demand of massive data and complex computation, network models are becoming more and more complex and contain more layers, which makes it much more difficult to implement them in hardware. Aiming at alleviating this issue, the proposed FPGA architecture provides reliable solutions through dynamic reconfiguration. Basically, part of the target network can be implemented in firstly, and then the rest of layers can be loaded without interruption by dynamic reconfiguration. In this way, those large network models can be successfully fit in a normal-size FPGA.

\newpage
\section*{\large Simulation Details of LUT}
Fig.~\ref{fig:supplementary_lut}(a) and (b) illustrate the simulation waveform of the select signal and the output signal during read stage in the 6-input FeFET LUT, respectively. All the simulations are done in HSPICE. As shown in Fig.~\ref{fig:supplementary_lut}(a), a pulse signal (1 V) is given to control the multiplexer and select LUT cells. During the read stage, different LUT cells would be selected and the configuration bits stored in would be passed to Output as Fig.~\ref{fig:supplementary_lut}(b) presents. According to the waveform and measurement, the average read delay is around 124.3 ps.

\begin{figurehere}
   \centering
    \includegraphics[scale=1,width=0.9\textwidth]{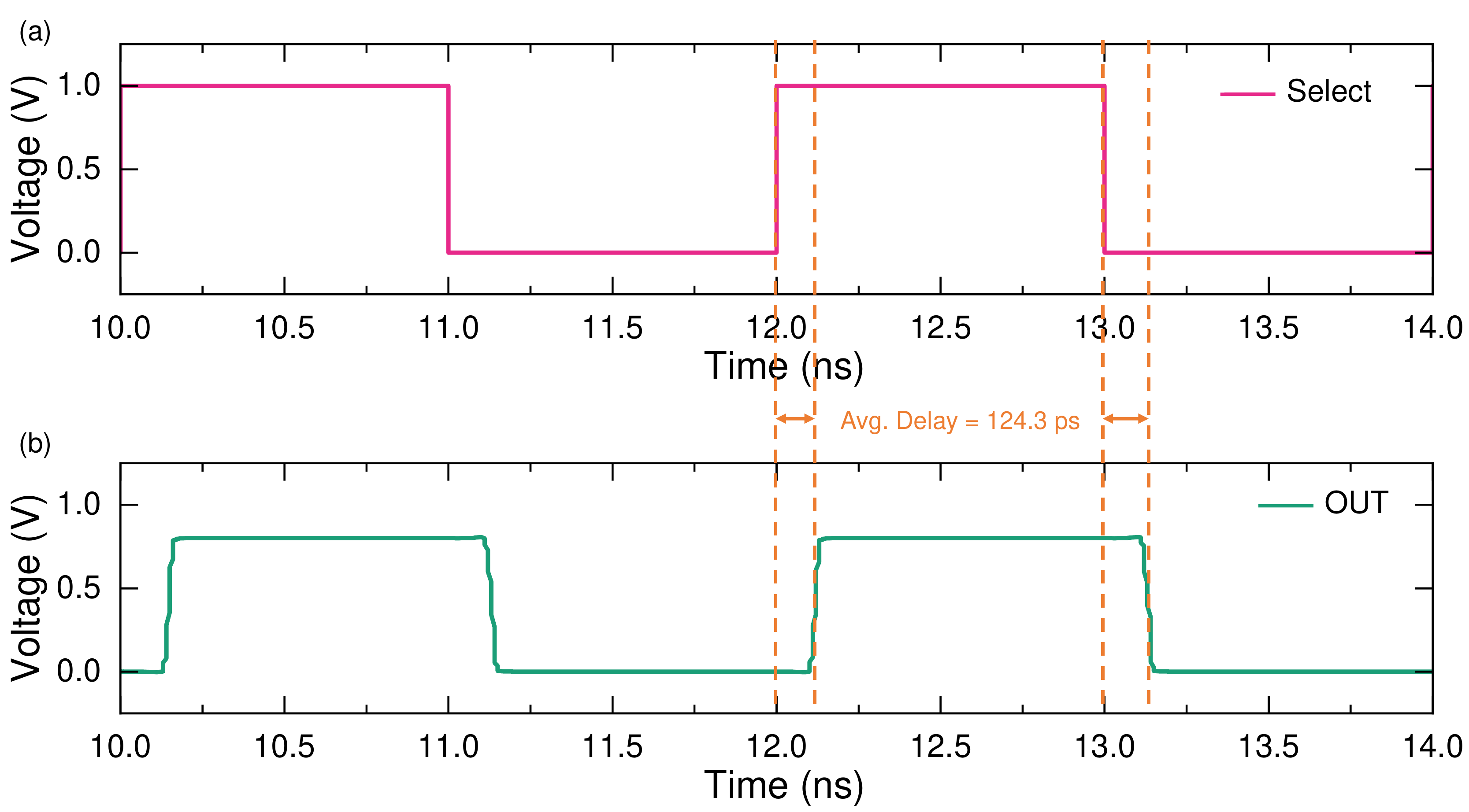}
    \captionsetup{parbox=none} 
    \caption{\textbf{Simulation waveform of the 6-input FeFET LUT.} (a) The simulation waveform of select signal in proposed 6-input LUT. (b) The simulation waveform of output signal in proposed 6-input LUT. The average read delay is around 124 ps.}
    \label{fig:supplementary_lut}
\end{figurehere}

\newpage
\section*{\large Two-Step Programming and Write Disturb of FeFETs}

Fig.\ref{fig:supplementary_write_disturb} illustrates the bias conditions for one configuration in the FeFET LUT during the second step of the two-step programming. After the first step, all the FeFETs have been programmed to the low-\textit{V}\textsubscript{TH} state. Then depending on the stored information, those FeFETs need to be at the high-\textit{V}\textsubscript{TH} state will be applied an -4 V across the gate and the body. For those FeFETs that need to stay at the low-\textit{V}\textsubscript{TH} state, inhibition biases are applied to the body such that the gate-to-body voltage drop is only -2V, not enough to disturb state. Such a scheme has been successfully verified in the experiment.

\begin{figurehere}
   \centering
    \includegraphics[scale=1,width=0.5\textwidth]{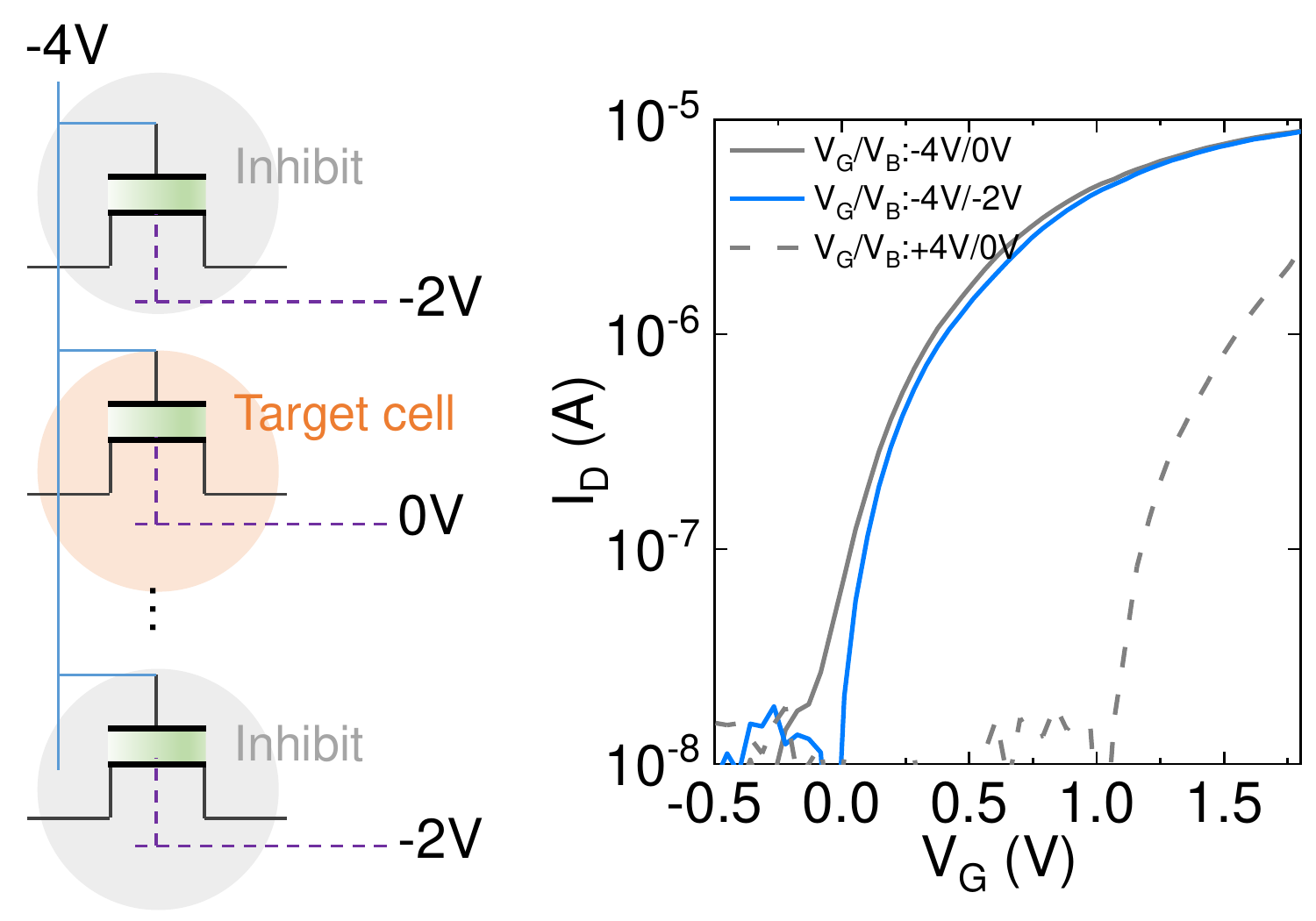}
    \captionsetup{parbox=none} 
    \caption{Illustrations of bias conditions for the second step of the two-step programming and the \textit{I}\textsubscript{D}-\textit{V}\textsubscript{G} characteristics of the low-\textit{V}\textsubscript{TH} and high-\textit{V}\textsubscript{TH} states and the half-selected cells (in blue). W/L=0.5$\mu$m/0.5$\mu$m.}
    \label{fig:supplementary_write_disturb}
\end{figurehere}

\newpage
\section*{\large Multi-Configuration CB Validation}

In addition to the one combination shown in the Fig.\ref{fig:CB}, where the branch 1/branch 2 is in the low-\textit{V}\textsubscript{TH}/high-\textit{V}\textsubscript{TH} states respectively, the other three combinations are also verified experimentally. Fig.\ref{fig:cb_hvt_hvt} shows the results when the branch 1/branch 2 are both in the high-\textit{V}\textsubscript{TH} states. In this case, no signal propagation happens, so the output remains low.

\begin{figurehere}
   \centering
    \includegraphics[scale=1,width=0.9\textwidth]{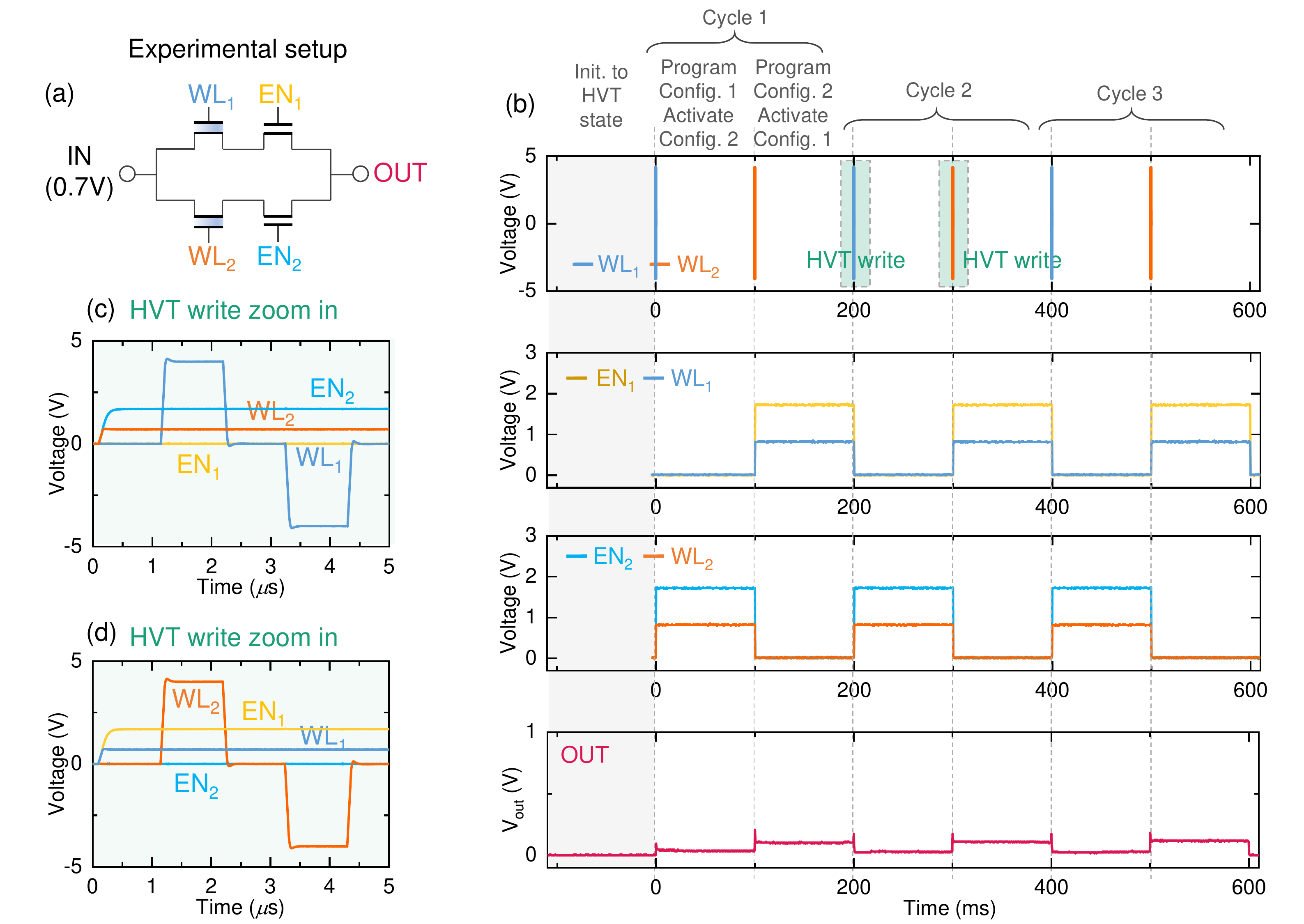}
    \captionsetup{parbox=none} 
    \caption{\textbf{Experimental verification of the multi-configuration CB  operation when both branches are in the high-\textit{V}\textsubscript{TH} states.} (a) The circuitry of one CB test unit. (b) The experimental transient waveforms of run-time context configuration and switching repeated for 3 cycles. (c)/(d) The zoomed-in programming waveform for branch 1/branch 2 to the high-\textit{V}\textsubscript{TH} state, respectively. }
    \label{fig:cb_hvt_hvt}
\end{figurehere}

Fig.\ref{fig:cb_lvt_lvt} shows the verification when the branch 1/branch 2 are both in the low-\textit{V}\textsubscript{TH} states. In this case, except after the initialization, the output should remain high due to the signal transmission, as also shown in the experimental results.

\begin{figurehere}
   \centering
    \includegraphics[scale=1,width=0.9\textwidth]{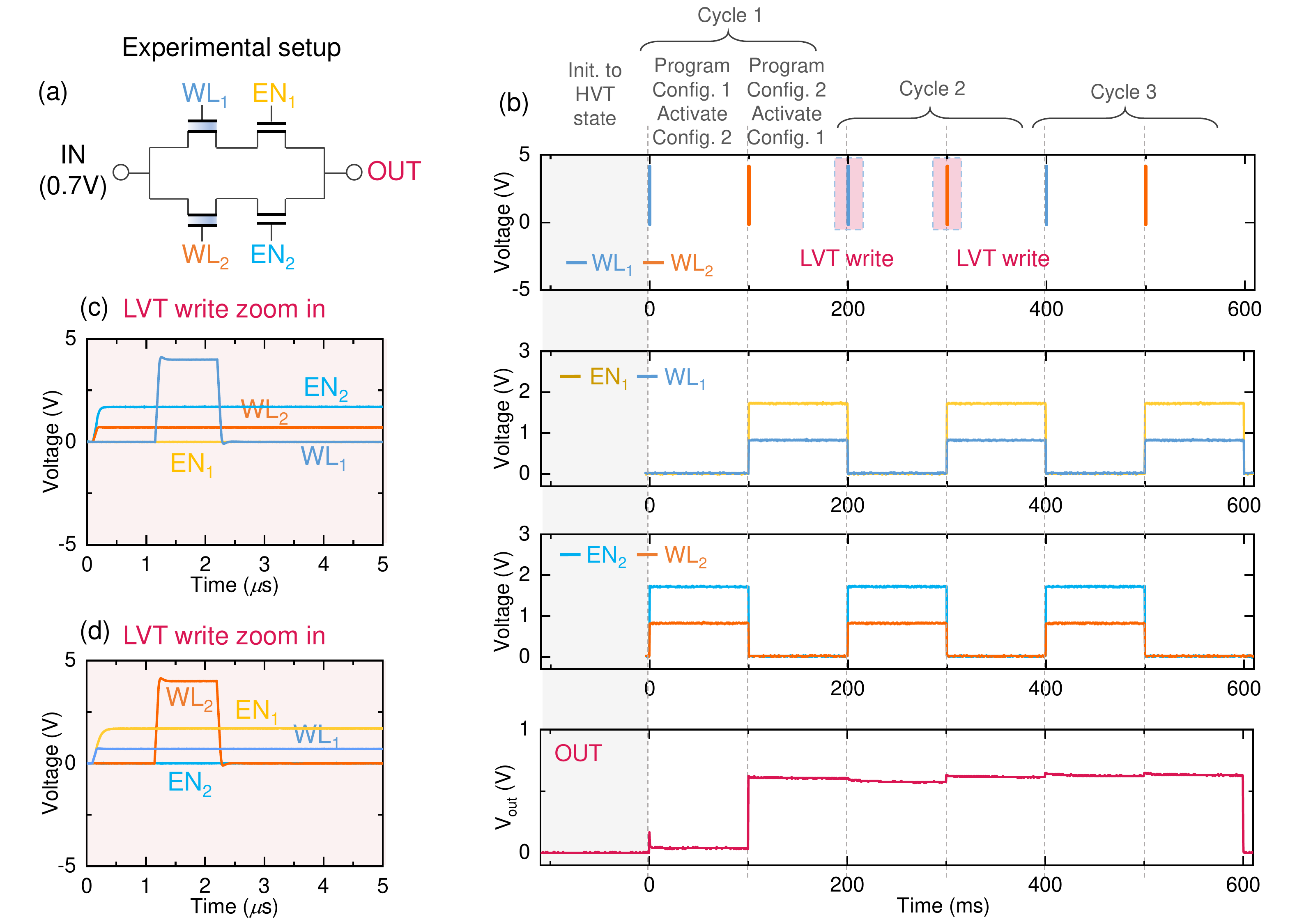}
    \captionsetup{parbox=none} 
    \caption{\textbf{Experimental verification of the multi-configuration CB operation when both branches are in the low-\textit{V}\textsubscript{TH} states.} (a) The circuitry of one CB test unit. (b) The experimental transient waveforms of run-time context configuration and switching repeated for 3 cycles. (c)/(d) The zoomed-in programming waveform for branch 1/branch 2 to the low-\textit{V}\textsubscript{TH} state, respectively. }
    \label{fig:cb_lvt_lvt}
\end{figurehere}

Fig.\ref{fig:cb_hvt_lvt} shows the verification when the branch 1/branch 2 are in the high-\textit{V}\textsubscript{TH}/low-\textit{V}\textsubscript{TH} states, respectively. In this case, the output will switch between high and low and it is high when the branch 2 is active. The first cycle is an exception because both branches are initialized to the high-\textit{V}\textsubscript{TH} states to begin with.

\begin{figurehere}
   \centering
    \includegraphics[scale=1,width=0.9\textwidth]{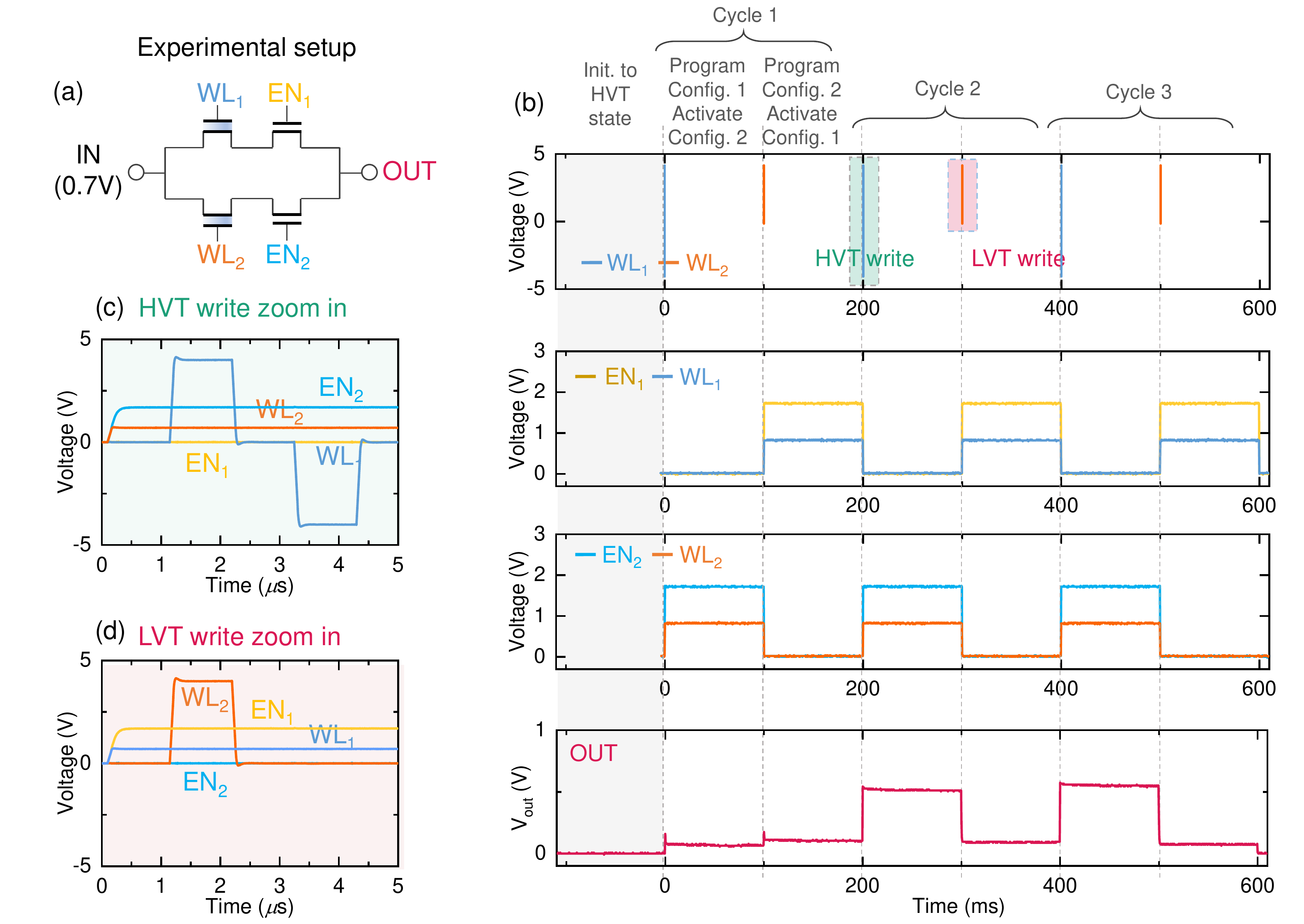}
    \captionsetup{parbox=none} 
    \caption{\textbf{Experimental verification of the multi-configuration CB  operation when branch 1/branch 2 are in the high-\textit{V}\textsubscript{TH}/low-\textit{V}\textsubscript{TH} states, respectively.} (a) The circuitry of one CB test unit. (b) The experimental transient waveforms of run-time context configuration and switching repeated for 3 cycles. (c)/(d) The zoomed-in programming waveform for branch 1/branch 2 to the high-\textit{V}\textsubscript{TH}/low-\textit{V}\textsubscript{TH} state, respectively. }
    \label{fig:cb_hvt_lvt}
\end{figurehere}

\newpage
\section*{\large Simulation Details of Multi-Configuration CB}
Fig.~\ref{fig:supplementary_cb} illustrates the simulation waveform of the input signal and the output signal in the multi-configuration FeFET CB, respectively. All the simulations are done in HSPICE. In the simulation, a pulse input signal (0.8 V) is asserted to pass through the FeFET CB (Fig.~\ref{fig:supplementary_cb}(a)). On the output terminal, the same pulse would be detected with the delay (Fig.~\ref{fig:supplementary_cb}(b)), which is around 7.8 ps on average.

\begin{figurehere}
   \centering
    \includegraphics[scale=1,width=0.9\textwidth]{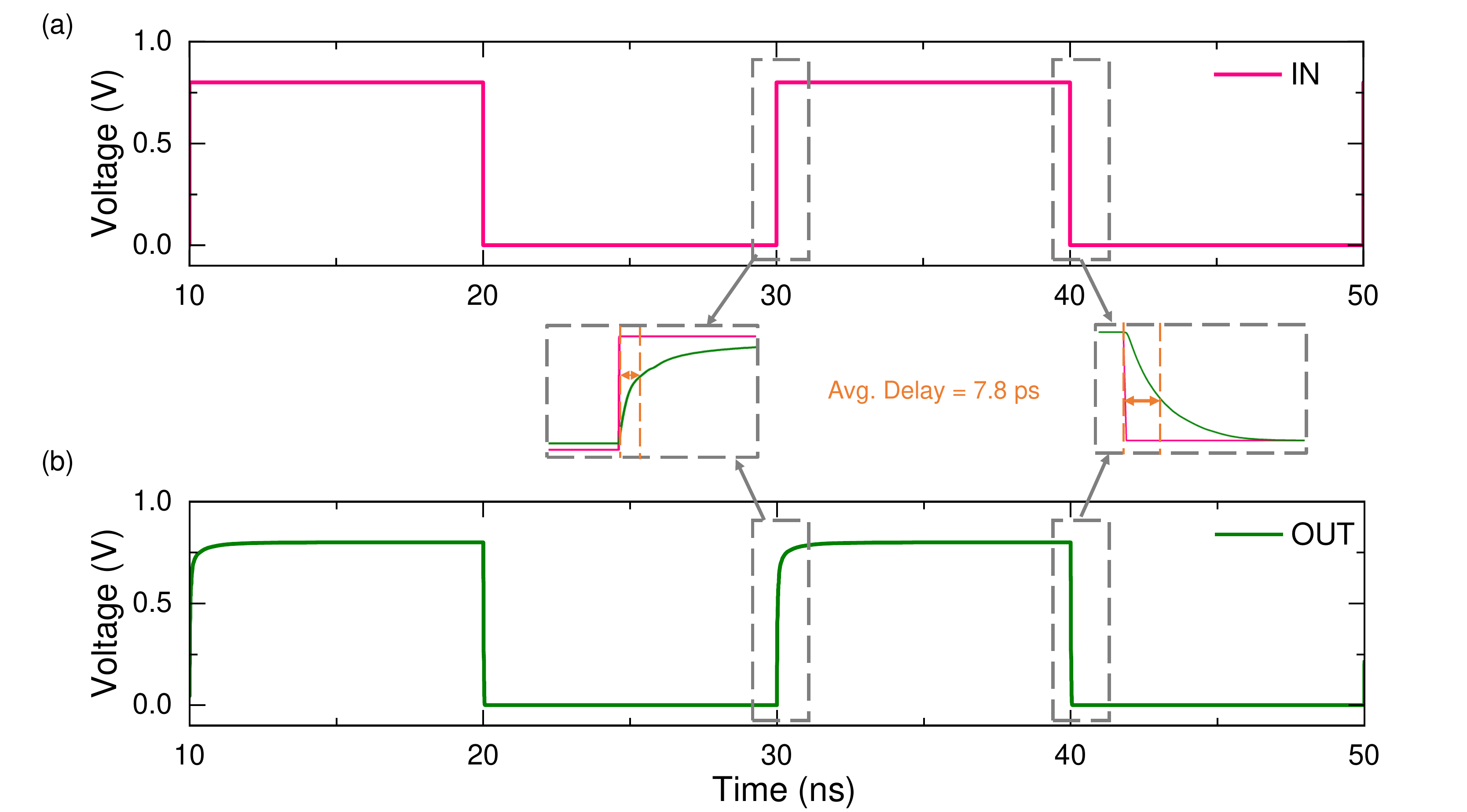}
    \captionsetup{parbox=none} 
    \caption{\textbf{Simulation waveform of the multi-configuration FeFET CB.} (a) The simulation waveform of input signal in proposed multi-configuration FeFET CB. (b) The simulation waveform of output signal in proposed multi-configuration FeFET CB. The average read delay is around 7.8 ps.}
    \label{fig:supplementary_cb}
\end{figurehere}

\newpage
\section*{\large Layout}
\begin{figure}[h]
   \centering
    \includegraphics[scale=1,width=1.0\textwidth]{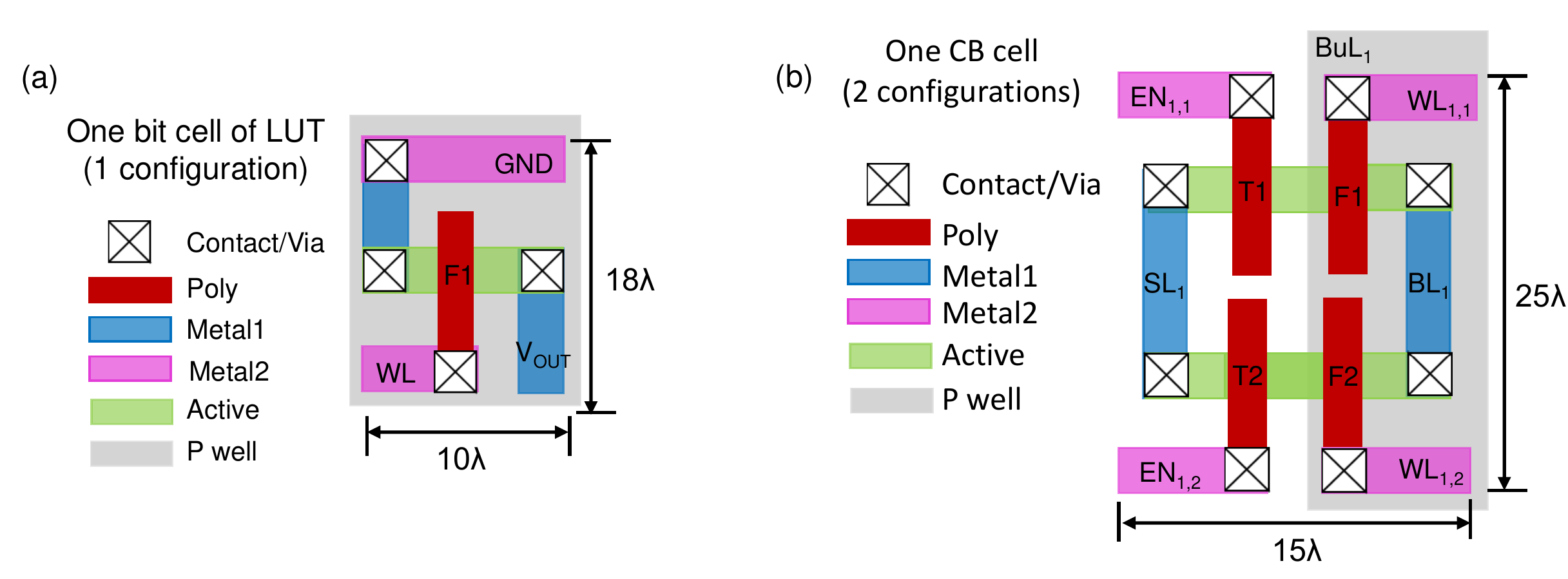}
    \captionsetup{parbox=none} 
    \caption{\textbf{Layout.} (a) Layout of a single LUT cell. (b) Layout of a cell of multi-configuration CB.}
    \label{fig:supplementary_layout}
\end{figure}

Fig.~\ref{fig:supplementary_layout}(a) shows the layout of a single LUT cell. The cell has 10 $\lambda$ width and 18 $\lambda$ length. The area is calculated as 180 $\lambda^{2}$. Fig.~\ref{fig:supplementary_layout}(b) shows the layout of a single cell of CB supporting dynamic reconfiguration. The cell has 15 $\lambda$ width and 25 $\lambda$ length. The area is calculated as 375 $\lambda^{2}$. Note, all the layouts follow the lambda design rules.

\newpage
\section*{\large Case Study}
In this section, we will introduce more cases of implementing our FPGA design into deep learning applications and show the benefits of our design.

The first case is regarding to dynamic configuration switching in DNN to show the performance improvement provided by dynamic reconfiguration in deep learning applications. Basically, there are two systems used in our case. As illustrated in Fig.~\ref{fig:supplementary-case_study}(a), in System 1, we deploy a Xilinx DPU B1152 core with softmax for accelerating an entire neural network. However, as a comparison, System 2 consists of a Xilinx DPU B2304 core without softmax for accelerating all but the last layer of a neural network, and a Xilinx DPU B1152 core with softmax for accelerating the last layer of the neural network and softmax layer. The simulation results show that System 2 which employs dynamic switching of layer resources yields more throughput ($\sim$1.7x) in DNN applications (Fig.~\ref{fig:supplementary-case_study}(b)). 

The other case is shown in Fig.~\ref{fig:supplementary-case_study}(c). Basically, in this case study we want to investigate the impact of dynamic reconfiguration on performance of FPGA in deep neural network domains. Hence, we implement 3 neural networks (ResNet50, CNV, and MobileNetv1) into Xilinx AIveo U250 card via Xilinx Vitis AI\cite{vitisai}. To get the reconfiguration time of each network, we use the formula that the size of the bitstream over the port throughput to calculate the reconfiguration time. And we assume the maximum bandwidth is performed with the reconfiguration ports (ICAP) which is 3.2 Gb/s\cite{bandwidth}. In addition, we run these built network models in Vitis AI and obtain the estimated latency reports which are the execution time of different networks in U250 board. In Fig.~\ref{fig:supplementary-case_study}(c), we do simulation for adding another condition as a supplementary of the case study shown in Fig.~\ref{fig:case_study}(c)\&(d) which is much more common in practical applications. In some applications performing multiple networks, we should firstly patch some of the networks before switching to another. The reason is that the former networks need to learn from these frames such that we can build a better network for current condition. In this situation, the feature of run-time reconfiguration of our design is able to serve these kinds of applications perfectly. In Fig.~\ref{fig:supplementary-case_study}(c), we show another time saving under the condition that executes the first network 5 times, then switch to the second one. The total time saving decreases a bit as it is expected, but still remains around 88.42\% at maximum. In conclusion, our architecture which offers the capability of dynamic reconfiguration provides significant benefit on latency for various deep learning applications.

\begin{figurehere}
   \centering
    \includegraphics[scale=1,width=1\textwidth]{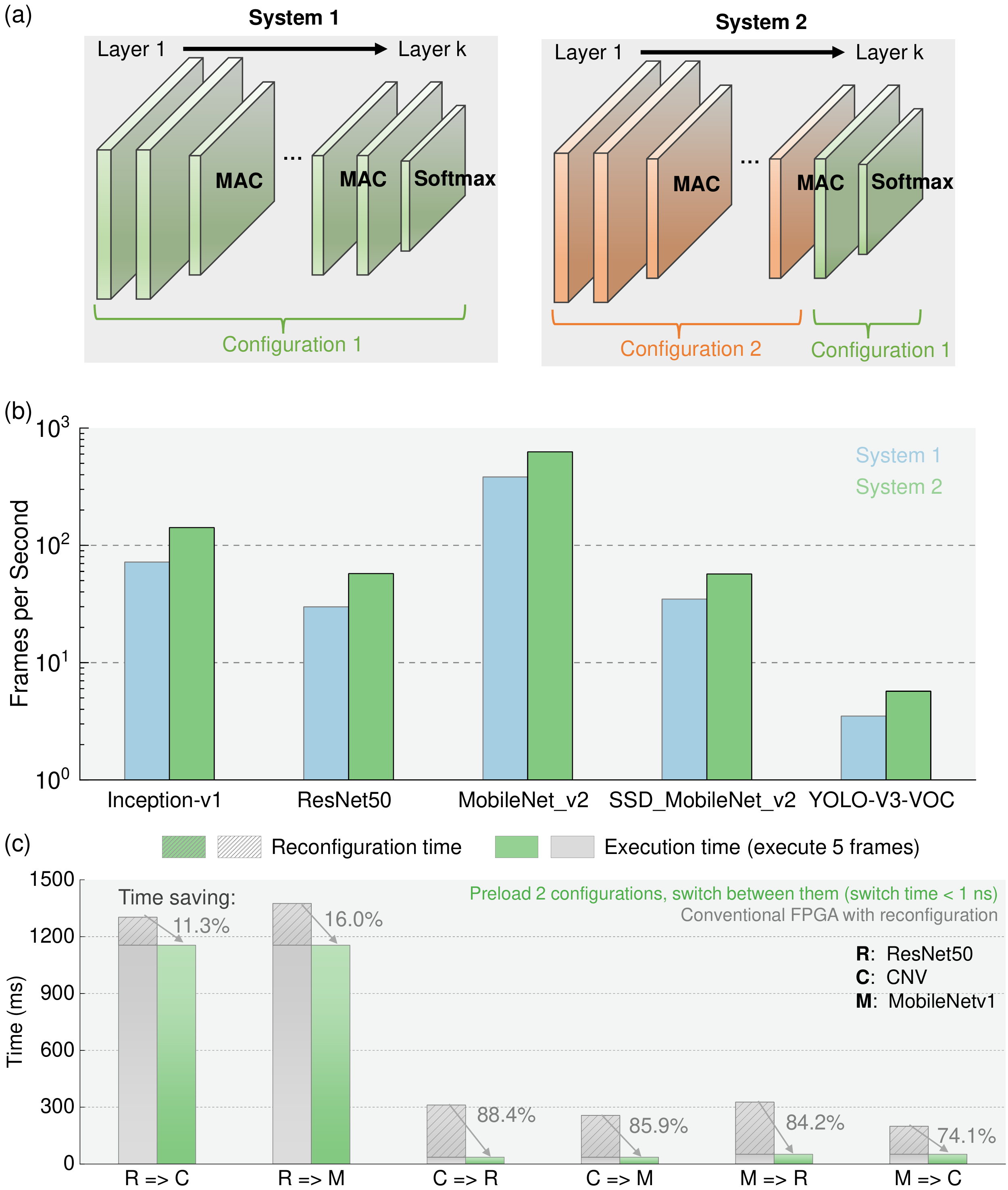}
    \captionsetup{parbox=none} 
    \caption{\textbf{Case study.} (a) In our case study of dynamically switching layer resources for DNN, 2 systems are used for showing the benefit brought by dynamically switching layer resources. System 1: a Xilinx DPU B1152 core with softmax for accelerating an entire neural network; System 2: a Xilinx DPU B2304 core without softmax for accelerating all but the last layer of a neural network, and a Xilinx DPU B1152 core with softmax for accelerating the last layer of the neural network and softmax layer. (b) Dynamic switching of layer resources in FPGA yields more throughput in DNN applications. (c) For some applications that need to be trained more, our design still shows significant time saving varying from 11.32\% to 88.42\%.}
    \label{fig:supplementary-case_study}
\end{figurehere}

\end{document}